\documentclass[twocolumn, tighten, times, twocolappendix]{aastex7}

\begin{document}

\title{Testing models for fully and partially stripped low-mass stars with \textit{Gaia}:\\ Implications for hot subdwarfs, binary RR Lyrae, and black hole impostors}

\author[orcid=0000-0002-1386-0603,gname=Pranav,sname=Nagarajan]{Pranav Nagarajan}
\affiliation{Department of Astronomy, California Institute of Technology, 1200 E. California Blvd., Pasadena, CA 91125, USA}
\email[show]{pnagaraj@caltech.edu}  

\author[orcid=0000-0002-6871-1752,gname=Kareem,sname=El-Badry]{Kareem El-Badry}
\affiliation{Department of Astronomy, California Institute of Technology, 1200 E. California Blvd., Pasadena, CA 91125, USA}
\email{kelbadry@caltech.edu}  

\author[orcid=0000-0002-4674-0704,gname=Alexey,sname=Bobrick]{Alexey Bobrick}
\affiliation{School of Physics and Astronomy, Monash University, Clayton, VIC 3800, Australia}
\affiliation{Australian Research Council Centre of Excellence for Gravitational Wave Discovery, Clayton, VIC 3800, Australia}
\email{alexey.bobrick@monash.edu} 

\author[orcid=0000-0003-0293-503X,gname=Giuliano,sname=Iorio]{Giuliano Iorio}
\affiliation{Institut de Ci\'encies del Cosmos, Universitat de Barcelona, Carrer de Martí i Franquès, 1, 08028 Barcelona, Spain}
\email{giuliano.iorio@icc.ub.edu}  

\author[orcid=0000-0003-3878-0498,gname=Francisco,sname=Molina]{Francisco Molina}
\affiliation{Institut für Physik und Astronomie, Universität Potsdam, Karl-Liebknecht-Str. 24/25, 14476, Golm, Germany}
\affiliation{Dpto. Química ``Prof. José Carlos Vílchez Martín'', Facultad de CC Experimentales, Universidad de Huelva, 21007 Huelva, Spain}
\email{francisco.molina@dqcm.uhu}  

\author[orcid=0000-0001-6172-1272,gname=Joris,sname=Vos]{Joris Vos}
\affiliation{Astronomical Institute of the Czech Academy of Sciences, CZ-25165, Ond\v{r}ejov, Czech Republic}
\email{joris.vos@uv.cl}

\author[orcid=0000-0001-8339-6423,gname=Maja,sname=Vu\v{c}kovi\'{c}]{Maja Vu\v{c}kovi\'{c}}
\affiliation{Instituto de Física y Astronomía, Universidad de Valparaíso, Gran Bretaña 1111, Playa Ancha, Valparaíso 2360102, Chile}
\email{maja.vuckovic@uv.cl}  

\begin{abstract}

When low-mass ($\lesssim 2\,M_{\odot}$) red giants lose their envelopes to a companion just before the helium flash, the resulting mass transfer can produce binaries hosting hot subdwarfs, horizontal branch stars, and undermassive red clump stars. Recent work predicts a continuum of such products, from fully stripped hot subdwarfs to partially stripped horizontal branch and red clump stars, and suggests that young, metal-rich RR Lyrae can form when partial stripping leaves a helium-burning star in the instability strip. To enable direct comparison with observations, we model these binaries in a simulated Milky Way–like galaxy with a realistic metallicity-dependent star formation history and 3D dust map, generate epoch astrometry using {\it Gaia}'s scanning law, and fit it with the cascade of astrometric models applied in {\it Gaia} DR3. We compare the simulated population to DR3 observations of hot subdwarfs, RR Lyrae, and red giants with high astrometric mass functions. The model significantly overpredicts the number of hot subdwarfs with astrometric binary solutions, partly because the predicted flux ratios are more unequal than observed. It also predicts $\gtrsim 100$ RR Lyrae with DR3 astrometric orbital solutions, while none are observed. We conclude that RR Lyrae in au-scale binaries may be substantially rarer than predicted. In contrast, the model plausibly explains the population of red clump stars with high astrometric mass functions, which we interpret as potential black hole impostors. We predict that $\sim10\times$ more stripped-star binaries will be detectable in DR4, whose sensitivity to longer periods will more strongly test wide-orbit systems.
\end{abstract}



\section{Introduction}
\label{sec:intro}

Binary interactions often result in the complete or partial stripping of a star's hydrogen-rich envelope \citep[e.g.,][]{han_origin_2002, han_origin_2003, heber_stripped_2016}. Depending on the degree of stripping, low-mass stripped stars (i.e., those with initial masses $\lesssim 2\,M_{\odot}$) that are currently undergoing core helium burning are predicted to be observable as hot subdwarf, horizontal branch, or red clump stars \citep[e.g.,][]{li_discovery_2022, vos_bobrick_2020, bobrick_iorio_2024}. These core helium-burning stars form a continuous population, with the fully stripped hot subdwarfs and the partially stripped red clump stars located at the blue (hot) and red (cool) ends of the horizontal branch on a color-magnitude diagram, respectively. Some partially stripped stars are predicted to pass through the instability strip and exhibit RR Lyrae-type pulsations while being younger and more metal-rich than classical RR Lyrae \citep{pietrzynski_rrl_2012, karczmarek_instability_2017, bobrick_iorio_2024}. 

Depending on whether the mass transfer between a red giant and its companion is stable or unstable, binary evolution is predicted to produce either a wide ($P_{\text{orb}} = 1$--$10$ yr), post-stable mass transfer binary or a close ($P_{\text{orb}} \lesssim 10$ d), post-common envelope binary \citep[e.g.,][]{paczynski_close_1972, hjellming_webbink_1987, soberman_phinney_1997, rodriguez_population_2025}. Wide binaries hosting fully or partially-stripped stars --- most of which are predicted to have main-sequence companions --- thus provide an opportunity to test models for stable mass transfer. The \textit{Gaia} mission has opened a new window on post-interaction binaries at intermediate orbital separations \citep{gaia_collaboration_gaia_2016, gaia_collaboration_gaia_2023, gaia_collaboration_gaia_2023-1, shahaf_wd_2024, yamaguchi_population_2025}, making it possible to study the population of au-scale binaries containing products of stable mass transfer in unprecedented detail. For example, \citet{bobrick_iorio_2024} predict $\approx 95\%$ of low-mass stripped stars to be found in binaries with orbital periods between $500$ and $2000$ days, a period range ideally suited for detection with \textit{Gaia} astrometry \citep[e.g.,][]{el-badry_gaias_2024}. 

\citet{vos_bobrick_2020} used detailed binary evolution modeling to perform population synthesis of low-mass (i.e., $M_{\text{primary, initial}} < 2.1\,M_{\odot}$) hot subdwarf (sdB) + main sequence binaries, taking into account the metallicity-dependent star formation history of the Milky Way. They were able to reproduce the observed orbital periods, mass ratios, metallicities, and formation rates of the population of such binaries observed at the time \citep{vos_sdb_2019}. \citet{bobrick_iorio_2024} used the same binary evolution models as \citet{vos_bobrick_2020} to construct a simulated Galactic population of binaries featuring various degrees of stripping. They were particularly motivated to investigate the properties of partially-stripped stars that fall in the instability strip and are observable as RR Lyrae variables. 

RR Lyrae produced by single-star evolution are traditionally only expected to be found in very old and metal-poor $(\mathrm{[Fe/H]} < -1)$ populations \citep[e.g.,][]{baade_nucleus_1946, preston_delta_1959, sandage_globular_1970}, having been used to study the Galactic bulge, Galactic halo, Magellanic Clouds, and dwarf galaxies in the Local Group \citep[e.g.,][]{pietrukowicz_deciphering_2015, murareva_smc_2018, cusano_vmc_2021, iorio_chemo_2021, nagarajan_rr_2022, savino_m31_2022}. In contrast, \citet{bobrick_iorio_2024} predict that RR Lyrae variables formed through binary interactions (i.e., via the formation channel proposed by \citealt{karczmarek_instability_2017}) are common, constituting the majority of the observed population of RR Lyrae that are young, metal-rich, and found on thin-disk Galactic orbits \citep[e.g.,][]{iorio_chemo_2021, dorazi_galah_2024, mateu_direct_2025, zhang_revealing_2025}. \citet{bobrick_iorio_2024} also predict that young and metal-rich RR Lyrae produced via partial envelope stripping by a binary companion should dominate the RR Lyrae population in the solar neighborhood, though more recent work has found that the metallicity threshold above which single-star evolution struggles to reproduce the observed properties of RR Lyrae is $\mathrm{[Fe/H]} \gtrsim -0.5$, somewhat higher than previously thought \citep{zhang_revealing_2025}. 

In this work, we apply the generative model of \citet{el-badry_generative_2024} to a simulated population of stripped star binaries in a Milky Way-like galaxy produced by the models of \citet{vos_bobrick_2020} and \citet{bobrick_iorio_2024} to predict the number of systems that are detectable in \textit{Gaia} DR3 and DR4. While \citet{vos_bobrick_2020} focused on hot subdwarfs and \citet{bobrick_iorio_2024} investigated RR Lyrae formed through the binary channel, this paper is aimed at understanding the intrinsic population of all low-mass stripped core helium-burning stars in our Galaxy. Comparing the predicted and observed populations allows us to further test their binary evolution model and assess the relative prevalence of partial and complete stripping \citep[e.g.,][]{xiong_blue_2022, li_discovery_2022}. Furthermore, by analyzing the stripped star binaries in the population of \citet{bobrick_iorio_2024} that have high astrometric mass functions, we also investigate the population of partially stripped, low-mass giants that can appear as black hole (BH) impostors \citep[e.g.,][]{bodensteiner_hr_2020, shenar_hidden_2020, el-badry_ngc_2022, el-badry_what_2022, el-badry_unicorns_2022}. A forthcoming companion paper will investigate the detectability of binarity in nearby RR Lyrae in more detail \citep{Giuliano26}.

The remainder of this paper is organized as follows. In Section~\ref{sec:modeling}, we mock observe the simulated population of \citet{bobrick_iorio_2024} and generate artificial catalogs of astrometric binary solutions. In Section~\ref{sec:results}, we compare the model's predictions against \textit{Gaia} DR3 observations of composite sdB binaries, RR Lyrae with non-single star solutions, and partially-stripped giants that can manifest as BH impostors. In Section~\ref{sec:discussion}, we discuss the implications of our results for population synthesis models of the Galactic population of stripped stars in binaries. Finally, in Section~\ref{sec:conclusion}, we summarize our key results and consider directions for future work.

\section{Population Synthesis} 
\label{sec:modeling}

\subsection{Galactic population of stripped stars}

\begin{figure*}
    \centering
    \includegraphics[width=\textwidth]{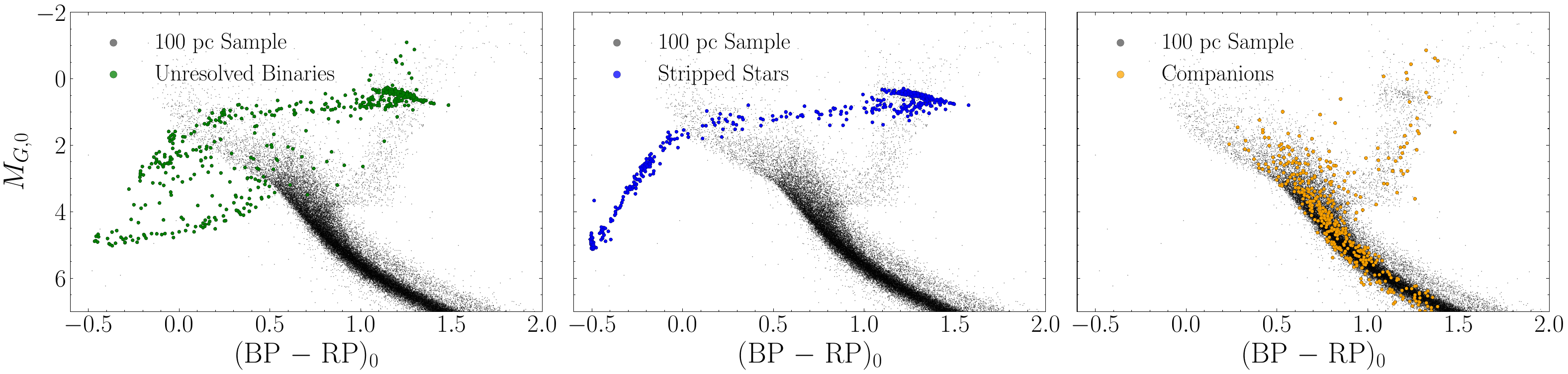}
    \caption{Extinction-corrected color-magnitude diagrams (CMDs) of both the simulated stripped star binaries and their individual components. We show a random sample of sources from \textit{Gaia} DR3 within 100 pc for comparison. The stripped stars lie in the red clump, horizontal branch, and hot subdwarf regions of the CMD. The light contributions of the luminous companions, which are MS stars and giants, add scatter to the CMD positions of the unresolved sources.}
    \label{fig:cmd}
\end{figure*}

\begin{figure*}
    \centering
    \includegraphics[width=\textwidth]{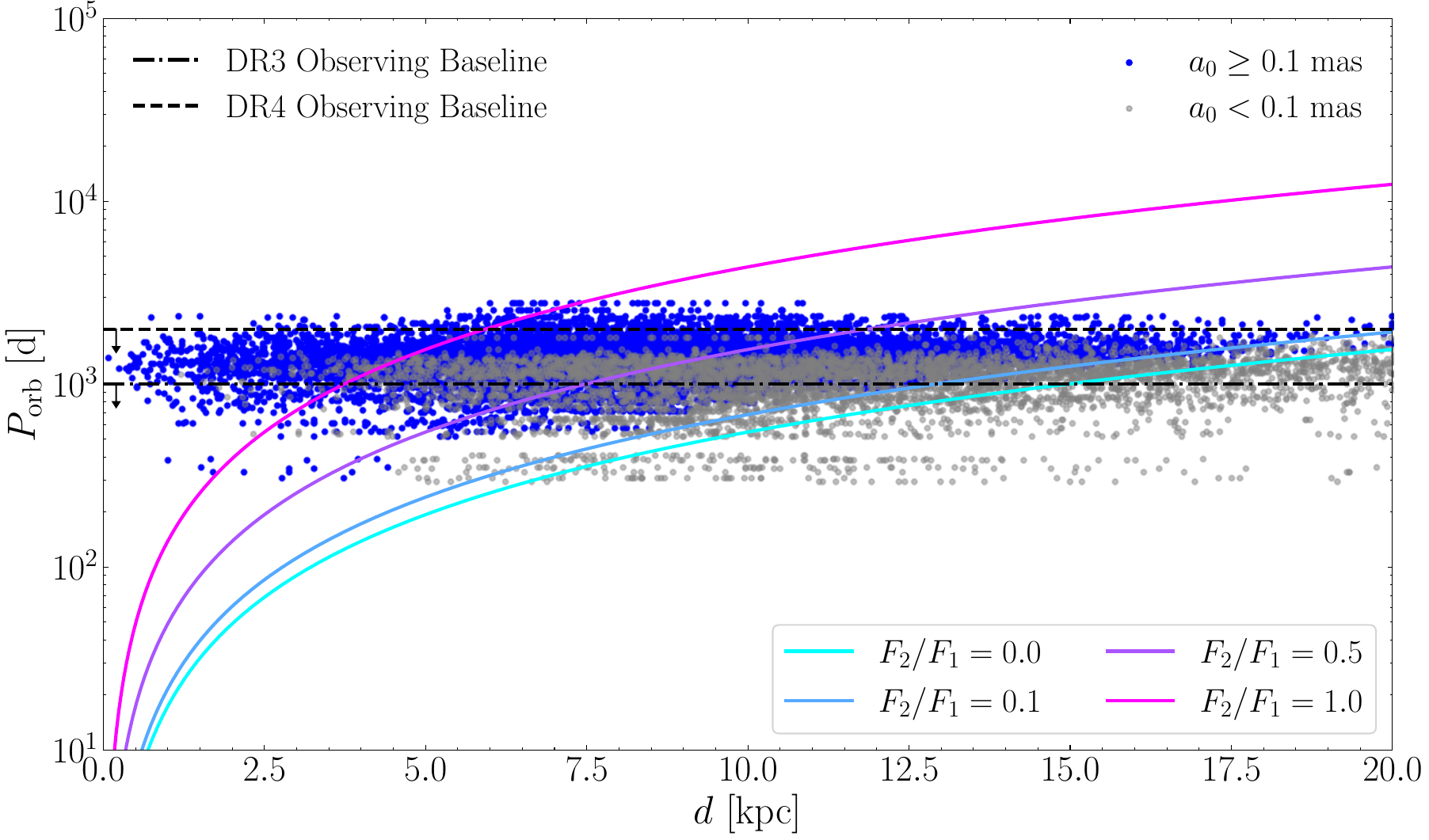}
    \caption{Orbital period vs.\ distance for the predicted population of stripped star binaries. For clarity, we show a random sample of 1\% of the binaries. We plot curves corresponding to $a_0 = 0.1$ mas (which is an approximate lower limit on the angular semi-major axis of detectable photocenter orbits in DR3) for different flux ratios $S \equiv F_2 / F_1$, assuming a typical stripped star mass of $0.5\,M_{\odot}$ and a typical main sequence star mass of $1\,M_{\odot}$. Binaries with $a_0 < 0.1$ mas are plotted in gray with lower opacity. We show the approximate upper detectable limit on the orbital period in DR3 ($\sim 1000$ d) with a dash-dotted line and the corresponding limit in DR4 ($\sim 2000$ d) with a dashed line.  The majority of stripped star binaries are amenable to detection via \textit{Gaia} astrometry.}
    \label{fig:sma}
\end{figure*}

A simulated population of stripped stars in binaries, described in \citet{vos_bobrick_2020} and \citet{bobrick_iorio_2024}, is derived based on detailed binary evolution modeling with \texttt{MESA} \citep[Modules for Experiments in Stellar Astrophysics,][]{paxton_2011, paxton_2013, paxton_2015, paxton_2018, paxton_2019, jermyn_2023}. The binaries are mapped onto the Besan\c{c}on model of the Galaxy, which describes the metallicity-dependent star formation history of the Milky Way and is calibrated by large photometric surveys \citep{robin_besancon_2003}. The masses of the primary (i.e., the initially more massive) stars are drawn from a Kroupa \& Haywood ``v6'' initial mass function \citep{haywood_vertical_1997, kroupa_imf_2008, czekaj_renewed_2014}, with a focus only on primaries that have masses between $0.7\,M_{\odot}$ and $2.1\,M_{\odot}$ and thus can ignite helium burning degenerately within the age of the Universe. The primaries are further restricted to those that become core helium-burning stars at the present time based on single-star evolutionary tracks from the MIST database \citep{choi_2016}.

The binaries' orbital periods are drawn from a log-uniform distribution \citep{abt_fraction_1983}, while the companion masses are drawn from a uniform distribution of secondary-to-primary mass ratios. Only binaries with initial orbital periods between $100$ and $700$ d and initial mass ratios between $0.\overline{33}$ and $1$ are simulated. These choices restrict the population to those systems that ignite helium degenerately and undergo stable mass transfer, as binaries with smaller mass ratios are expected to undergo common envelope evolution \citep[e.g.,][]{han_origin_2002, han_origin_2003}. The close (i.e., orbital periods
between $1$--$10^4$ d) binary fractions in the field and the halo are assumed to be 0.25 and 0.40, respectively \citep{moe_close_2019}. The mass transfer is assumed to be conservative until the accretor achieves super-critical rotation, at which point the mass transfer becomes fully non-conservative \citep[][]{popham_narayan_1991, paczynski_polytropic_1991, deschamps_critical_2013}. Since red giants typically tidally circularize by the time they begin Roche-lobe overflow \citep[e.g.,][]{vos_eccentricity_2015}, the simulated binaries are assumed to have zero eccentricity.

The Neural Network-assisted Population Synthesis code NNaPS \citep{nnaps} is used to predict observable properties from the \texttt{MESA} runs. Based on their effective temperatures, the stripped stars are classified as B-type hot subdwarfs (sdBs, with $T_{\text{eff}} > 20,000$ K, formed by complete stripping), ``A-type hot subdwarfs'' (sdAs, with $20,000$ K $> T_{\text{eff}} > 15,000$ K, formed by near-complete stripping), or horizontal branch stars (HB stars, with $T_{\text{eff}} < 15,000$ K, formed by partial stripping). Horizontal branch stars that fall within the instability strip defined by \citet{karczmarek_instability_2017} are identified as RR Lyrae (RRL). For further details on initial conditions, assumed binary physics, and extraction of observational properties, we refer the reader to \citet{vos_bobrick_2020} and \citet{bobrick_iorio_2024}. In this work, Galactic positions for the binaries are drawn using an adjusted version of the Galactic population script from the LISA Synthetic UCB Catalogs project (Bobrick et al., in prep.; Breivik et al., in prep.).\footnote{https://github.com/Synthetic-UCB-Catalogs/Galaxy-scripts}

We plot an extinction-corrected color-magnitude diagram (CMD) of the stripped star binaries in the left panel of Figure~\ref{fig:cmd}. We estimate the individual $\mathrm{BP}-\mathrm{RP}$ colors of the stripped star and main sequence components of these binaries from their effective temperatures using MIST \citep{choi_2016}, and show them along with the individual $G$-band magnitudes of those components in the center and right panels of Figure~\ref{fig:cmd}, respectively. A random sample of \textit{Gaia} DR3 sources within 100 pc is shown for reference. For these sources, we retrieve reddenings from the 3D dust map of \citet{green_2019}, and  calculate extinctions in the $G$, BP, and RP bands based on the $R_V = 3.1$ extinction law of \citet{cardelli_1989}.  The simulated unresolved binaries are predicted to fall across the red clump, horizontal branch, and hot subdwarf regions of the CMD, with some scatter due to the contribution of the luminous secondary. 

We plot the orbital periods of the stripped star binaries versus their distances in Figure~\ref{fig:sma}, selecting a random sample of 1\% of the simulated population for clarity. We also plot curves corresponding to $a_0 = 0.1$ mas (which is the approximate lower detectable limit on the angular semi-major axis of the photocenter in DR3; e.g., \citealt{el-badry_generative_2024}) for different values of the secondary-to-primary $G$-band flux ratio $S \equiv F_2 / F_1$. In doing so, we adopt typical values for the stripped star mass and main sequence star mass of $0.5\,M_{\odot}$ and $1\,M_{\odot}$, respectively. We do not assume specific evolutionary states for the components, reflecting the fact that flux ratio can vary for a given mass ratio. We plot systems with $a_0 < 0.1$ mas in gray with lower opacity. We show the approximate upper detectable limits on orbital period in DR3 ($\sim 1000$ days) and DR4 ($\sim 2000$ days) with dash-dotted and dashed lines, respectively. Many binaries that visually fall within the marked detectability limits have $a_0 < 0.1$ mas (and are not actually detectable in DR3) because they have component masses and photocenter semi-major axes that are significantly different from the values assumed for the plotted curves in Figure~\ref{fig:sma}. Nevertheless, Figure~\ref{fig:sma} shows that the majority of stripped star binaries have $a_0 > 0.1$ mas and $P_{\text{orb}} < 2000$ d and are thus amenable to detection via \textit{Gaia} astrometry.

\subsection{Mock observations}

\begin{figure*}
    \centering
    \includegraphics[width=\textwidth]{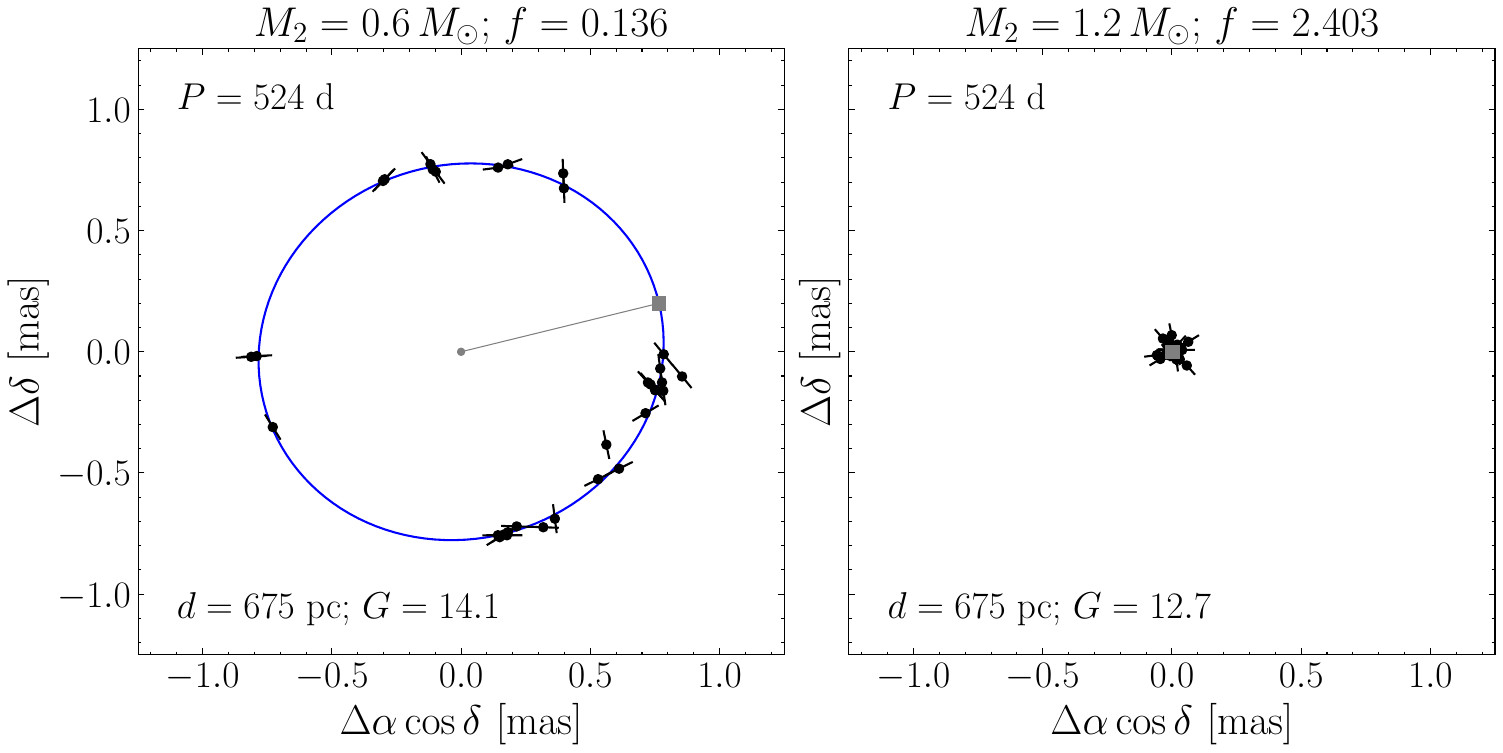}
    \caption{Mock DR3 observations of two unresolved $\approx 0.47\,M_{\odot}$ sdB binaries. In the left panel, the binary hosts an $\approx 0.6\,M_{\odot}$ main sequence companion, leading to a small secondary-to-primary flux ratio of $\approx 0.136$. In the right panel, the companion mass is doubled to $\approx 1.2\,M_{\odot}$ with all other parameters held constant, leading to a larger flux ratio of $\approx 2.403$. Due to the larger photocenter orbit, the \texttt{gaiamock} pipeline predicts that only the binary on the left receives an astrometric binary solution in DR3, underscoring the fact that detectability depends significantly on flux ratio.}
    \label{fig:mock_orbits}
\end{figure*}

We follow the approach of \citet{nagarajan_realistic_2025} to simulate the \textit{Gaia} observations and astrometric model fitting of stripped star binaries. Specifically, we bootstrap 9 additional populations by choosing different random locations of the solar neighborhood in the simulated galaxy at a radius of 8.1 kpc (i.e., by rotating about the $z$-axis), resulting in 10 total realizations. While this effectively modifies the orientation of the central bar, no binaries at those distances are predicted to receive orbital solutions in DR3 or DR4 (see Figure~\ref{fig:mw_dist_figure}), so this lack of rotational symmetry does not affect our results. We model extinction for each source using the \texttt{Combined19} 3D dust map in the \texttt{mwdust} package \citep{bovy_2016}, which combines the maps of \citet{drimmel_2003}, \citet{marshall_2006}, and \citet{green_2019}. Then, assuming that $A_G = 2.8 E(B - V)$, where $E(B - V)$ is the color excess on the scale of \citet{schlegel_1998}, we calculate extinctions $A_G$ and apparent magnitudes $m_G$ in the \textit{Gaia} $G$-band. We assign a random orientation to each binary in each realization, sampling the cosine of the inclination, the argument of periastron, the longitude of the ascending node, and the periastron time from uniform distributions. The tightest orbits published in DR3 have $a_0 \approx 0.2$ mas, and all sources with orbital solutions have apparent magnitude $m_G < 19$. To speed up our computations, we adopt preliminary cuts of $a_0 > 0.1$ mas, $m_G < 19$, and $d < 5$ kpc or $10$ kpc in DR3 or DR4, respectively.

We pass the synthetic population of stripped star binaries through the \texttt{gaiamock}\footnote{\texttt{https://github.com/kareemelbadry/gaiamock}} pipeline \citep{el-badry_generative_2024} to determine whether or not each system receives an astrometric binary solution in each \textit{Gaia} data release. For a given binary with specified parameters, the \texttt{gaiamock} pipeline generates mock observations using the \textit{Gaia} scanning law and fits the epoch astrometry using \textit{Gaia}'s astrometric model cascade to determine what kind of solution (i.e., single-star, acceleration, or orbital) the binary will receive. We consider systems in the generated astrometric binary catalog to be ``detected'' if they pass the detectability cuts in \textit{Gaia} DR3 (see \citealt{halbwachs_gaia_2023} for details). 

We show mock DR3 observations of two unresolved sdB binaries in Figure~\ref{fig:mock_orbits}. In both binaries, the mass of the stripped star is $\approx 0.47\,M_{\odot}$, the orbital period is $524$ d, and the distance to the system is $675$ pc. However, in the left panel, the binary hosts an $\approx 0.6\,M_{\odot}$ main sequence companion, leading to a small secondary-to-primary flux ratio of $\approx 0.136$. In the right panel, the companion mass is doubled to $\approx 1.2\,M_{\odot}$, leading to a larger flux ratio of $\approx 2.403$. This combination of flux ratio and mass ratio results in a photocenter that is nearly coincident with the binary's center-of-mass, so the binary in the left panel has a much larger photocenter orbit. Consequently, the \texttt{gaiamock} pipeline predicts that only the binary in the left panel receives an astrometric binary solution in DR3, while the one on the right would receive a single-star solution with little astrometric excess noise. Figure~\ref{fig:mock_orbits} demonstrates that the detectability of binaries with astrometry depends sensitively on flux ratio.

\section{Results}
\label{sec:results}

\subsection{Hot subdwarfs in binaries}

\begin{deluxetable*}{ccccccc}
\tablecaption{All known and candidate hot subdwarf binaries with astrometric binary solutions in DR3. The reported parallaxes are from the non-single star catalog. \label{tab:sdB_sols}}
\tablehead{\colhead{\textit{Gaia} DR3 ID} & \colhead{RA} & \colhead{Dec} & \colhead{$G$} & \colhead{$\varpi$} & \colhead{$P_{\mathrm{orb}}$} & \colhead{Solution Type} \\
& $\circ$ & $\circ$ & & mas & days & \\
\colhead{(1)} & \colhead{(2)} & \colhead{(3)} & \colhead{(4)} & \colhead{(5)} & \colhead{(6)} & \colhead{(7)}}
\startdata
Gaia DR3 392046852459641472 & 4.835820 & 46.551884 & 14.245 & $0.638 \pm 0.023$ & 884.105 & Orbital \\
Gaia DR3 972725503164737152 & 90.400987 & 53.482230 & 13.932 & $1.444 \pm 0.052$ & 854.190 & Orbital \\
Gaia DR3 1191985885728851328 & 241.022876 & 14.413173 & 14.377 & $0.965 \pm 0.027$ & --- & Acceleration7 \\
Gaia DR3 1806581377789928448 & 301.914950 & 13.678857 & 14.111 & $0.917 \pm 0.031$ & 846.758 & Orbital \\
Gaia DR3 1975687674885816576 & 331.785597 & 47.790715 & 14.568 & $0.585 \pm 0.021$ & 759.041 & Orbital \\
Gaia DR3 2183496162809214464 & 307.729601 & 54.387065 & 15.126 & $0.667 \pm 0.023$ & 1137.907 & Orbital \\
Gaia DR3 2670789408307249024 & 324.236282 & -6.281383 & 13.631 & $0.885 \pm 0.022$ & --- & Acceleration7 \\
Gaia DR3 3540092300847749760 & 175.212699 & -22.912812 & 13.153 & $1.199 \pm 0.019$ & 617.173 & Orbital \\
Gaia DR3 3649963989549165440 & 218.378596 & -1.245270 & 14.300 & $1.367 \pm 0.051$ & 892.531 & Orbital \\
Gaia DR3 4477210283154292480 & 277.356266 & 6.784572 & 14.846 & $1.091 \pm 0.038$ & 732.364 & Orbital \\
Gaia DR3 4494329271178956928 & 263.572182 & 12.866425 & 13.502 & $1.268 \pm 0.018$ & --- & Acceleration7 \\
Gaia DR3 4522995326025050496 & 275.083902 & 17.158817 & 14.673 & $0.811 \pm 0.029$ & 977.601 & Orbital \\
Gaia DR3 4525830416761444096 & 279.256017 & 21.220840 & 14.586 & $1.010 \pm 0.035$ & 949.516 & Orbital \\
Gaia DR3 4550114402362108416 & 261.951565 & 16.749090 & 13.475 & $1.192 \pm 0.024$ & 821.436 & Orbital \\
Gaia DR3 4555610723553451392 & 264.690085 & 20.697343 & 14.712 & $0.744 \pm 0.022$ & --- & Acceleration7 \\
Gaia DR3 4723810851270083072 & 48.875408 & -59.568031 & 13.313 & $1.005 \pm 0.016$ & --- & Acceleration7 \\
Gaia DR3 4935975569903358848 & 30.393260 & -53.728752 & 13.013 & $1.071 \pm 0.014$ & --- & Acceleration7 \\
Gaia DR3 5579436712515286016 & 104.340833 & -35.193645 & 13.151 & $2.268 \pm 0.041$ & 751.074 & Orbital \\
Gaia DR3 5769521481016936960 & 187.591082 & -85.235635 & 13.828 & $1.220 \pm 0.025$ & 658.923 & Orbital \\
Gaia DR3 6335746093599431296 & 225.526755 & -4.993986 & 13.353 & $1.205 \pm 0.033$ & 717.942 & Orbital \\
Gaia DR3 6359368722966483840 & 270.509015 & -83.285258 & 14.057 & $1.008 \pm 0.021$ & 740.078 & Orbital \\
Gaia DR3 6587335519633433472 & 331.450852 & -35.327445 & 13.176 & $1.273 \pm 0.025$ & 828.138 & Orbital \\
Gaia DR3 6631822855308840320 & 283.263122 & -61.435955 & 14.572 & $0.933 \pm 0.028$ & 634.328 & Orbital \\
\enddata
\end{deluxetable*}

\begin{figure*}
    \centering
    \includegraphics[width=\textwidth]{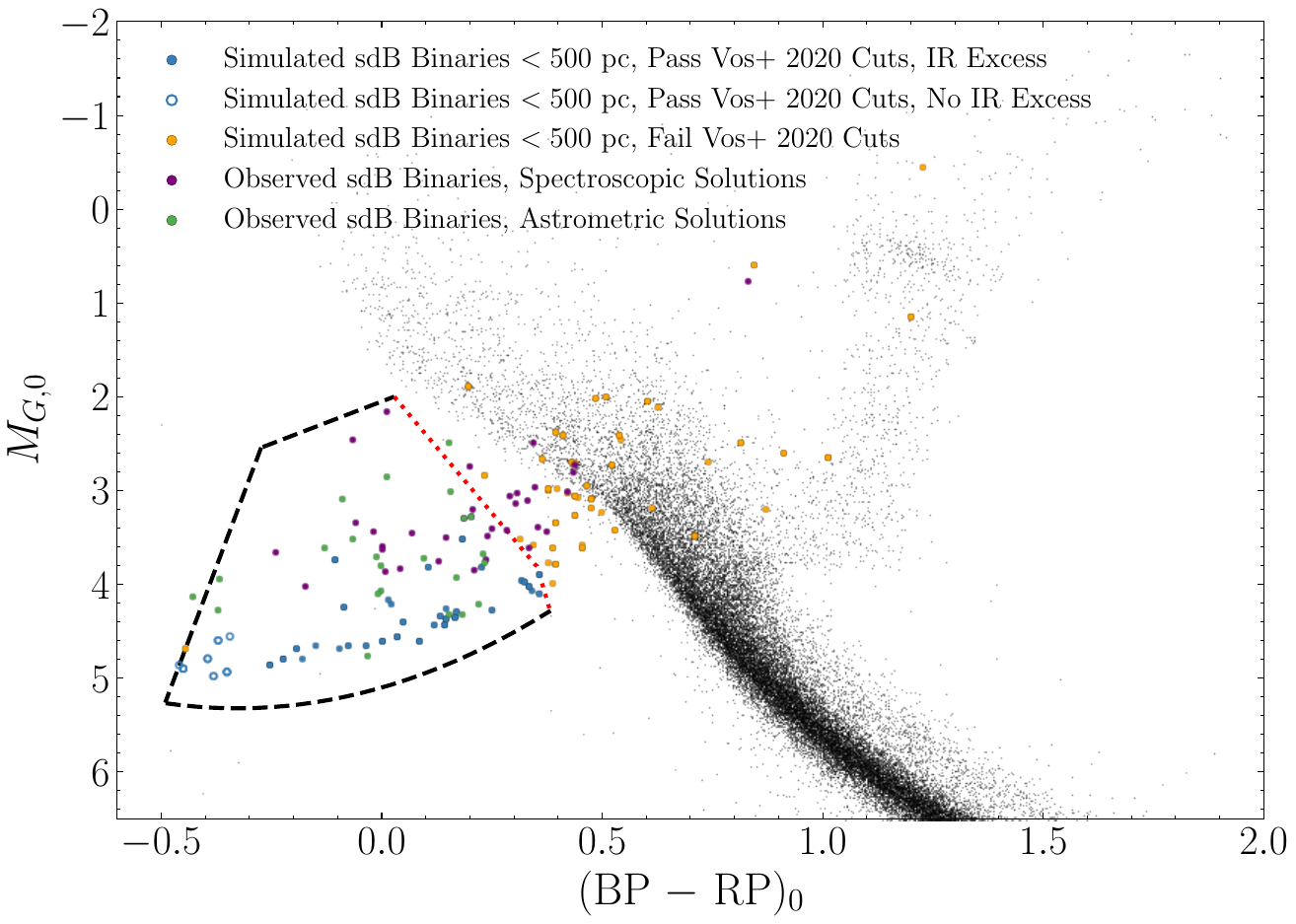}
    \caption{Simulated sdB binaries within 500 pc in the \textit{Gaia} color-magnitude diagram (CMD). Many of these binaries are duplicates located at different Galactic coordinates. For comparison, we also show the known wide composite sdB binaries with published astrometric \citep{geier_dr2_2019, culpan_subdwarf_2022} or spectroscopic \citep{barlow_two_2013, molina_wide_2026} orbital solutions. The majority of the observed systems pass the cuts used by \citet{vos_bobrick_2020} (black dashed lines) to identify composite sdB binaries. On the other hand, many simulated sdB binaries fall outside the red, dotted line, which is used by \citet{vos_bobrick_2020} (following \citealt{geier_dr2_2019}) to  exclude main sequence stars. It appears that this color cut is rather conservative, since several known composite sdBs with published orbits fall redward of it. We use open circles to denote simulated sdB binaries that are not predicted to show infrared excess, and hence would not be observationally identified as composite \citep[e.g.,][]{nemeth_hot_2012}. These systems lie toward the blue end of the CMD.}
    \label{fig:vos_cmd}
\end{figure*}

\begin{figure*}
    \centering
\includegraphics[width=\textwidth]{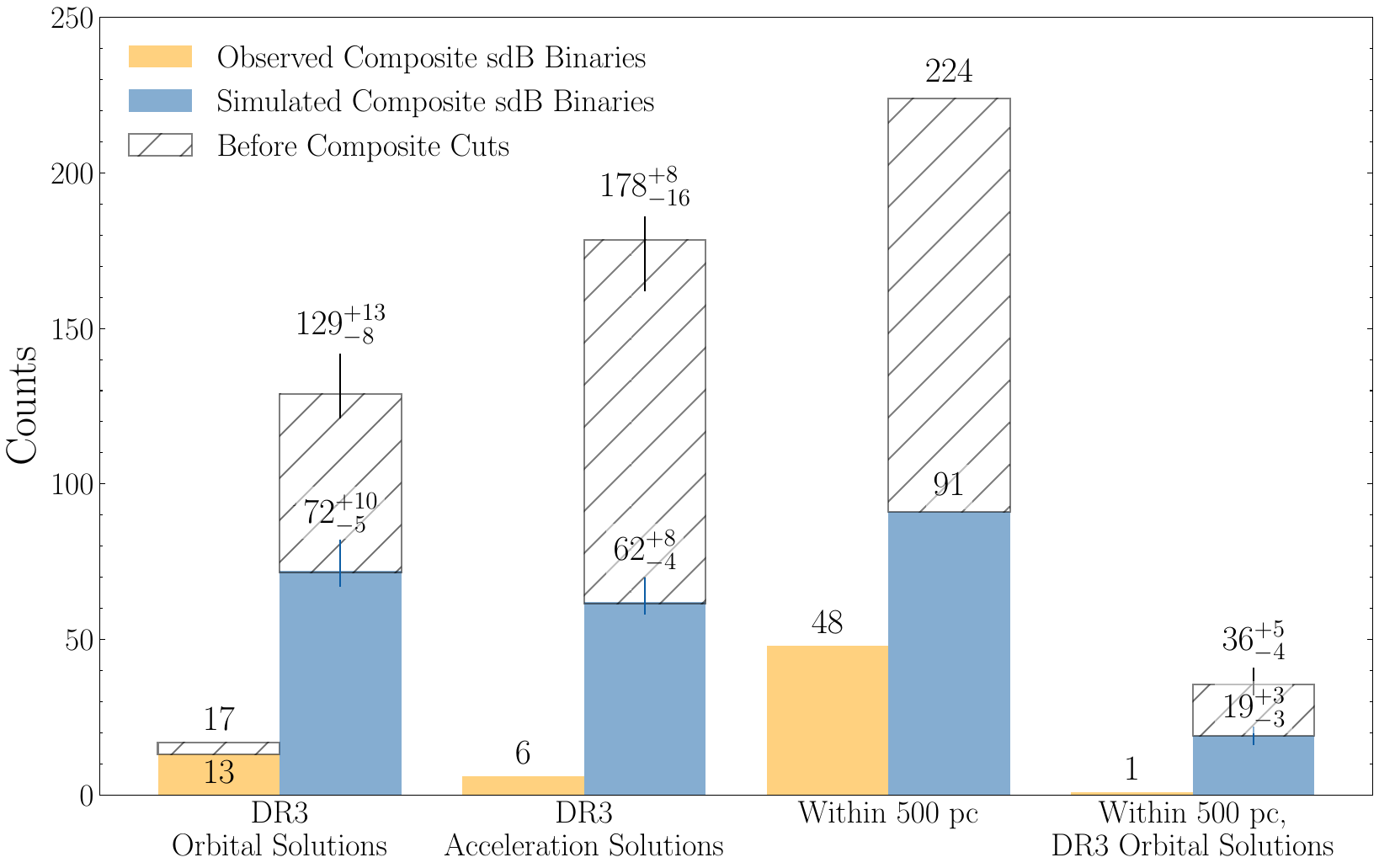}
    \caption{Median number of sdBs in composite binaries predicted to receive astrometric binary solutions in DR3 across 10 realizations. The error bars signify the middle 68\% of counts. The hatched bars show the number of sdB binaries in each category before the cuts used to identify composite systems are applied. We compare predictions against observations, finding that the adopted model significantly overestimates the number of composite sdB systems that receive astrometric binary solutions.}
    \label{fig:obs_bar_plot}
\end{figure*}

\begin{figure*}
    \centering
    \includegraphics[width=\textwidth]{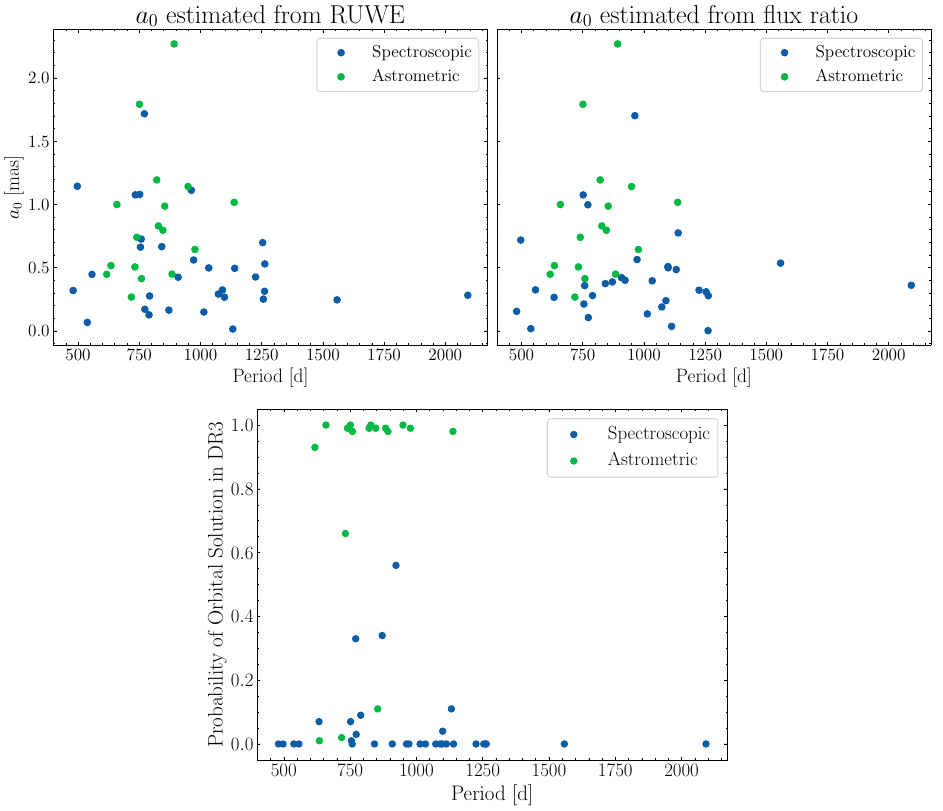}
    \caption{Comparison of wide-orbit sdB binaries with astrometric orbits from \textit{Gaia} (green; cross-matched with the samples of \citealt{geier_dr2_2019} and \citealt{culpan_subdwarf_2022}) and those with spectroscopic orbits from radial velocity monitoring (blue; \citealt{barlow_two_2013, molina_wide_2026}). In the top left panel, we estimate $a_0$ for each spectroscopic binary from the RUWE in DR3, while in the top right panel, we derive $a_0$ from published constraints on flux ratios in the literature. DR3 astrometry is most sensitive to orbits with large $a_0$ and $P_{\text{orb}} \lesssim 1000$ d. At $P_{\text{orb}} < 1000$ d, most sdB binaries with $a_0 \gtrsim 1$ mas received orbital solutions. At smaller $a_0$, some systems received orbital solutions and others did not, with the difference owing to the details of the \textit{Gaia} scanning law and quality cuts imposed on the solutions published in DR3. In the bottom panel, we show the \texttt{gaiamock}-estimated probability for each system to receive an orbital solution in DR3, restricting the sample of spectroscopic binaries to those with flux ratio constraints in the literature. All of the spectroscopic binaries have detection probabilities $\lesssim 0.5$, with the majority having detection probabilities $\lesssim 0.1$. On the other hand, almost all the detected astrometric binaries have detection probabilities near unity. Three of the astrometric binaries have low estimated detection probabilities (see text).}
    \label{fig:a0_comp_plot}
\end{figure*}

\subsubsection{Hot subdwarfs with astrometric binary solutions in DR3}

\citet{culpan_subdwarf_2022} compiled a catalog of 6,616 known hot subdwarfs with spectroscopic or photometric classifications and 61,585 candidates in (E)DR3. Cross-matching all known hot subdwarfs and hot subdwarf candidates with the DR3 non-single star catalog reveals that 12 of these binaries have astrometric orbital solutions, while 2 binaries have 7-parameter acceleration solutions. \citet{culpan_subdwarf_2022} built upon the work of \citet{geier_dr2_2019}, who identified 39,800 hot subluminous star candidates in DR2. Compared to \citet{geier_dr2_2019}, \citet{culpan_subdwarf_2022} applied more stringent selection cuts to account for potential crowded region effects. Indeed, cross-matching all hot subluminous star candidates from \citet{geier_dr2_2019} with the DR3 non-single star catalog returns 17 binaries with astrometric orbital solutions and 6 binaries with 7-parameter acceleration solutions, representing a proper superset of the cross-matched binaries from \citet{culpan_subdwarf_2022}. We check archival survey imaging of each candidate astrometric binary in the catalog of \citet{geier_dr2_2019} that is not present in the catalog of \citet{culpan_subdwarf_2022}, confirming that crowding is not a concern in each case. Hence, we consider all hot subdwarf candidates with astrometric solutions from either the \citet{geier_dr2_2019} or the \citet{culpan_subdwarf_2022} catalogs. We list these known and candidate hot subdwarfs with astrometric binary solutions in Table \ref{tab:sdB_sols}. 


The candidate list in Table~\ref{tab:sdB_sols} is not complete and has a poorly characterized selection function. However, \citet{dawson_volume_2024} construct a 500 pc volume-limited sample of confirmed hot subdwarfs that they estimate to be 90\% complete. They discover 48 systems that show an infrared excess in their spectral energy distribution (SED) fits, as expected for a composite sdB binary formed by stable mass transfer. Only one of these systems, Gaia DR3 5579436712515286016, has an astrometric orbital solution in the DR3 non-single star catalog. None of these systems have acceleration solutions. 

\subsubsection{Composite sdB binaries in the \textit{Gaia} color-magnitude diagram}

The \citet{bobrick_iorio_2024} model predicts that $129^{+13}_{-8}$ sdBs and $178^{+8}_{-16}$ sdBs with main sequence or red giant companions would have received orbital solutions and 7-parameter acceleration solutions in DR3, respectively. Not all these systems would have been identifiable as containing sdBs based on their colors and magnitudes. Most known long-period sdB binaries are ``composite'' systems with detectable flux contributions from both the sdB and the companion. \citet{vos_bobrick_2020} propose a series of cuts in the \textit{Gaia} color-magnitude diagram (Equations 5--9 in their paper) that identify composite sdB binaries while excluding main sequence contaminants. We show these cuts, along with all simulated sdB binaries within 500 pc (of which many are duplicates located at different Galactic coordinates), in Figure~\ref{fig:vos_cmd}. For comparison, we also show the known wide composite sdB binaries with published astrometric \citep{geier_dr2_2019, culpan_subdwarf_2022} or spectroscopic \citep{barlow_two_2013, molina_wide_2026} orbital solutions. While the majority of the observed systems pass the cuts used by \citet{vos_bobrick_2020} to identify composite sdB binaries, many simulated systems fall outside the color cut used by \citet{vos_bobrick_2020} and \citet{geier_dr2_2019} to exclude main sequence contaminants (shown with a red dotted line). This cut appears to be somewhat conservative, since several known composite sdBs with published orbits fall redward of it, as previously noted in the literature \citep{culpan_subdwarf_2022, dawson_volume_2024}.

sdBs with companions that are sufficiently faint may not be recognized as composite systems. We use the color cuts from \citet{nemeth_hot_2012} to select sdBs that can be identified as composite via infrared excess, requiring $V - J > 0$ and $J - H > 0$. To predict colors in these bands for the simulated sdB binaries, we retrieve template spectra for the companions from the BaSeL stellar library \citep{lejeune_1997, lejeune_1998} based on their modeled masses, radii, and effective temperatures. We retrieve model spectra for the sdBs from \citet{gotberg_unifying_2018}, interpolating in effective temperature and rescaling to match the simulated radii. For each binary, we integrate the total flux over the corresponding wavelength ranges of the Johnson $V$, 2MASS $J$, and 2MASS $H$ filters, predict $V - J$ and $J - H$ colors, and check if these satisfy $V - J > 0$ and $J - H > 0$. We use open circles in Figure~\ref{fig:vos_cmd} to denote simulated sdB binaries that are not predicted to show infrared excess. These systems, which would not be identified as composite, lie toward the blue end of the CMD. Of the observed binaries with astrometric orbital solutions, there are three that that are very blue and do not show any infrared excess. These systems have large astrometric mass functions and thus likely have white dwarf or neutron star companions \citep[e.g.][]{Geier2023}, making them different from any of the simulated systems or the previously known long-period sdBs with radial velocity (RV) orbits.

Applying the \citet{vos_bobrick_2020} and \citet{nemeth_hot_2012} cuts, we find that $17 \pm 2$ composite sdB binaries within 500 pc are predicted to receive orbital solutions, representing $\approx 19\%$ of the 91 composite sdB binaries in the simulated population within that distance limit.\footnote{If we apply the \citet{culpan_subdwarf_2022} criteria on top of the \citet{vos_bobrick_2020} and \citet{nemeth_hot_2012} cuts, then we find that 74 simulated sdB binaries within 500 pc would be identified as composite systems.} Of all sdBs in the simulated population, $\approx43$\% are predicted to pass the composite cuts. Of all sdBs in the simulated population within 500 pc, $\approx41$\% are predicted to pass the composite cuts. Most of the observed sdBs with astrometric or spectroscopic binary solutions fall within the cuts, but $\approx34$\% fall just outside of them. The observed systems are systematically brighter than the simulated ones, which might imply that the sdB masses are larger than predicted. 

We compare the median number of composite sdB binaries predicted to receive astrometric orbital solutions in DR3 across 10 realizations to observations in Figure~\ref{fig:obs_bar_plot}. The error bars signify the middle 68\% of counts. The hatched bars show the number of sdB binaries in each category before the composite cuts are applied. Based on comparison to DR3 observations, we find that the model overestimates the number of composite sdB binaries that receive orbital solutions by a factor of $\approx 5.5$ and the number of composite sdB binaries that receive 7-parameter acceleration solutions by a factor of $\approx 10$. However, we emphasize that the observed samples are not complete (i.e., there may be unrecognized sdBs with astrometric solutions). The model also overestimates the number of sdBs in composite binaries within 500 pc -- which is likely nearly complete --  by a factor of $\approx 2$, and the number of composite sdBs binaries with astrometric orbital solutions within 500 pc by a factor of $\approx 19$. We discuss potential explanations for these discrepancies in Section~\ref{sec:discussion}.


\subsubsection{Comparison of sdBs in astrometric and spectroscopic binaries}
\label{sec:comparison}

No sdBs with composite spectra and published RV orbits received astrometric orbital solutions in DR3. 
To understand why, we compare the orbital periods and photocenter semi-major axes of the observed samples of wide-orbit sdB binaries with astrometric orbits from \textit{Gaia} (i.e., cross-matched with the catalogs of \citealt{geier_dr2_2019} and \citealt{culpan_subdwarf_2022}) and those with spectroscopic orbits from radial velocity monitoring \citep{barlow_two_2013, molina_wide_2026} in Figure~\ref{fig:a0_comp_plot}. There is no overlap between the two samples, and no sdBs with spectroscopic orbits have acceleration solutions in DR3.

While $a_0$ is measured by \textit{Gaia} for the astrometric binaries, it must be derived for the spectroscopic binaries. In the top left panel of Figure~\ref{fig:a0_comp_plot}, we use Equation~1 of \citet{nagarajan_spectroscopic_2025} to estimate $a_0$ from the RUWE in DR3 \citep{el-badry_generative_2024}. On the other hand, in the top right panel, we derive $a_0$ based on published constraints on flux ratios in the literature \citep{barlow_two_2013, molina_wide_2026}.\footnote{The sdB + K0 system GALEX J022836.7-362543 appears to have an underestimated flux ratio estimate in the literature (see e.g., \citealt{vos_sdb_2019}) for its color. Using appropriate template spectra, we predict flux ratio as a function of extinction-corrected color. We then adopt a revised $G$-band flux ratio of $\approx 0.9$ based on its measured color of $BP-RP = 0.27$ \citep{gaia_collaboration_gaia_2023} and 3D dust map extinction of $E(B-V) = 0.02$ \citep{wang_dust_2025}.} In detail, we retrieve model spectra for the stripped stars from \citet{gotberg_unifying_2018}, choosing the model corresponding to an sdB mass of $0.51\,M_{\odot}$, consistent with the canonical value of $0.47 \pm 0.05\,M_{\odot}$ \citep[e.g.,][]{vos_orbits_2017}.\footnote{The sdB progenitors modeled by \citet{gotberg_unifying_2018} do not ignite helium degenerately. However, when estimating the flux ratio, the sdB's effective temperature is its most relevant parameter. The model we adopt corresponds to $T_{\text{eff}} = 28.2$ kK, reasonable for a typical sdB.} For the companions, we assume $\log g = 4.0$ (typical for a F/G-type dwarf) and solar metallicity, and retrieve the appropriate BOSZ \citep{bohlin_2017, meszaros_2024} template spectrum for the effective temperature corresponding to the companion's estimated spectral type. For each binary with an estimated flux ratio in the literature, we integrate over the corresponding wavelength range and rescale the template spectra to match the published constraint. Finally, we integrate over the $G$-band to determine the flux ratio in that filter, and use the published mass ratios, semi-major axes, and parallaxes to compute the photocenter semi-major axis of each spectroscopic binary. The two methods of estimating $a_0$ lead to roughly similar results, but with some system-to-system scatter that is likely mainly a result of poorly estimated flux ratios.


From Figure~\ref{fig:a0_comp_plot}, we observe that DR3 astrometry is most sensitive to orbits with large $a_0$ and $P_{\text{orb}} \lesssim 1000$ d. At $P_{\text{orb}} < 1000$ d, most sdB binaries with $a_0 \gtrsim 1$ mas received orbital solutions, while at smaller $a_0$, some systems received orbital solutions and others did not. The difference lies in the details of the \textit{Gaia} scanning law and quality cuts imposed on the solutions published in DR3; for instance, the non-linear DR3 cut on \texttt{parallax\_over\_error} penalizes more distant systems at fixed $a_0$, hindering these systems from receiving orbital solutions. 

To test this quantitatively, we retrieve the DR3 sky coordinates, parallaxes, and $G$-band magnitudes for each sdB binary with a published spectroscopic solution. We also retrieve their measured mass ratios and orbital parameters from the literature. To estimate the component masses, we assume the canonical sdB mass of $0.47\,M_{\odot}$. Adopting our estimated $G$-band flux ratios, we simulate 100 random binary orientations, and predict the probability of each spectroscopic binary receiving an orbital solution in DR3. We show these detection probabilities in the bottom panel of Figure~\ref{fig:a0_comp_plot}. We find that all detection probabilities are $\lesssim 0.5$, with most probabilities $\lesssim 0.1$. It is thus unsurprising that the composite sdB + main sequence binaries with spectroscopic orbital solutions in the literature did not receive astrometric orbits in DR3, mostly reflecting the fact that they have large flux ratios and thus small photocenter orbits. Indeed, from Figure~\ref{fig:vos_cmd}, we see that the sdBs with astrometric solutions are systematically bluer than those with spectroscopic solutions, because the former are biased toward low flux ratios and the latter toward high flux ratios. 

To confirm this, we repeat the exercise for the observed sdBs with astrometric orbital solutions in DR3. We retrieve their sky coordinates, parallaxes, $G$-band magnitudes, and orbital parameters (i.e., period, eccentricity, periastron time, and Thiele-Innes coefficients) from DR3, and predict the probability of each binary receiving an orbital solution in that data release, varying the random seed across each of 100 trials. We show these detection probabilities in the bottom panel of Figure~\ref{fig:a0_comp_plot} as well. As expected, we find that almost all of these binaries have detection probabilities near unity. Three of the astrometric binaries have low predicted detection probabilities, since they are predicted to narrowly fail the DR3 cut on parallax significance as a function of period. The fact that these systems nevertheless received orbits in DR3 suggests that \texttt{gaiamock} underestimates the scatter in parallax errors at fixed magnitude.

\subsubsection{Summary of sdB comparison}
The \citet{bobrick_iorio_2024} model reproduces the number of composite sdB binaries within 500 pc to within a factor of $\sim 2$. However, the model significantly overestimates the number of sdB binaries that receive DR3 astrometric solutions. Specifically, the model overestimates the number of composite sdB binaries that receive orbital solutions by a factor of $\approx 5.5$, the number of composite sdB binaries that receive 7-parameter acceleration solutions by a factor of $\approx 10$, and the number of composite sdB binaries within 500 pc that receive orbital solutions by a factor of $\approx 19$. 

We use \texttt{gaiamock} to understand why no sdBs with composite spectra and published RV orbits received astrometric orbital solutions in DR3. We find that these systems have low astrometric detection probabilities in DR3, in contrast to the observed sample of sdBs in astrometric binaries; we infer that this is due to the observed sample of sdBs in spectroscopic binaries being biased toward large flux ratios and small photocenter orbits. The most important tension between the model and observations is thus that the model predicts the existence of a large number of wide-orbit sdBs with relatively low-mass companions ($\lesssim 0.7\,M_{\odot}$) and large flux ratios, which are not observed. Indeed, in the adopted model, $\approx 11\%$, $\approx 23\%$, and $\approx 31\%$ of sdB binaries have companion masses $< 0.6\,M_{\odot}$, $< 0.7\,M_{\odot}$, and $< 0.8\,M_{\odot}$, respectively. Such systems are not easily detectable as composite systems in spectroscopic searches, but they are detectable with astrometry, and the DR3 data shows that they are significantly less common than predicted.

\subsection{sdA stars}

\citet{bobrick_iorio_2024} define ``A-type" hot subdwarfs, or ``sdA stars,'' as near-completely stripped stars with effective temperatures between 15,000 K and 20,000 K. However, A-type dwarfs have effective temperatures between 7,400 K and 10,000 K. The term ``sdA star'' was originally coined by \citet{kepler_sdA_2016} to refer to all sources with narrow hydrogen line spectra, $6.5 > \log g \geq 5.5$, and $T_{\text{eff}} \leq 20,000$ K. Both observational \citep[e.g.,][]{hermes_sdA_2017, brown_physical_2017, pelisoli_problem_2018, pelisoli_sdA_2018, pelisoli_sdA_2019} and theoretical \citep[e.g.,][]{yu_formation_2019} follow-up studies have concluded that the vast majority of these sources are metal-poor main sequence stars, with a small minority being extremely low-mass white dwarfs. 

While near-completely stripped stars should still show blue or UV excess, almost all follow-up of hot subdwarf candidates has found $T_{\text{eff}} \gtrsim 20,000$ K \citep{geier_dr2_2019, culpan_subdwarf_2022}. \citet{dawson_volume_2024} identified four objects which they classified as sdAs in their 500 pc volume-limited sample of hot subluminous stars. These objects are cooler and less luminous than sdBs, and thus clearly separated from that population of hot subdwarfs on a \textit{Gaia} CMD. They acknowledged that sdA-type spectra are characteristic of many different evolutionary classes, and that these objects require further follow-up. 

\citet{dawson_volume_2024} also identified  eight sources that lie in the same region of the CMD as the ``sdA stars'' in the simulated population of \citet{bobrick_iorio_2024}. However, they classified these objects as blue horizontal branch (BHB) stars instead. Recent work has found an overall intrinsic binarity fraction of these core helium-burning stars of just $\approx 30\%$, rising to $\approx 50\%$ for BHB stars on ``disk-like'' orbits \citep{guo_bhb_2025}. An even more recent study limited the binary fraction of BHB stars to $< 2.2\%$ \citep{culpan_bhb_2025}, and remarked that the work of \citet{guo_bhb_2025} may have been affected by contamination. It is possible that some of the observed BHBs are the observational analogs of what \citet{bobrick_iorio_2024} identify as sdAs. However, their model classifies 6\% of the simulated Galactic population, or 96,000 binaries, as ``sdA stars,'' with $131^{+8}_{-14}$ of these objects predicted to receive orbital solutions in DR3. This appears to be significantly larger than the population that is actually observed.


\subsection{RR Lyrae in binaries}

\begin{deluxetable*}{ccccccc}
\tablecaption{All candidates that are listed in both the \textit{Gaia} RR Lyrae and non-single star catalogs. The SB1 and EclipsingBinary solutions are spurious, as the observed spectroscopic or photometric variability likely arises from stellar pulsations. We do not identify any RR Lyrae in binaries. \label{tab:rrl_binaries}}
\tablehead{\colhead{\textit{Gaia} DR3 ID} & \colhead{RA} & \colhead{Dec} & \colhead{$G$} & \colhead{$P_{\mathrm{orb}}$} & \colhead{Solution Type} & \colhead{Reason for Rejection} \\
& $\circ$ & $\circ$ & & days & & \\
\colhead{(1)} & \colhead{(2)} & \colhead{(3)} & \colhead{(4)} & \colhead{(5)} & \colhead{(6)} & \colhead{(7)}}
\startdata
Gaia DR3 1193137104465445888 & 239.413379 & 16.000215 & 11.610 & 0.280 & SB1 & Spurious \\
Gaia DR3 3669003373413914624 & 218.922995 & 5.069565 & 14.633 & 0.608 & EclipsingBinary & Spurious \\
Gaia DR3 4073896567545411328 & 283.382925 & -25.770458 & 15.252 & 0.662 & EclipsingBinary & Spurious \\
Gaia DR3 4207618935502749184 & 290.898546 & -7.413276 & 15.498 & 0.800 & EclipsingBinary & Spurious \\
Gaia DR3 4285576371518811776 & 280.594692 & 6.245459 & 18.049 & 0.686 & EclipsingBinary & Spurious  \\
Gaia DR3 491104466350124544 & 51.004460 & 64.691708 & 14.065 & --- & Acceleration9 & Misclassified \\
Gaia DR3 5239771349915805952 & 160.213824 & -64.713236 & 12.014 & 2288 & Orbital & Misclassified \\
Gaia DR3 5764934181067920256 & 217.263761 & -88.645463 & 12.249 & 0.670 & SB1  & Spurious \\
Gaia DR3 6032843799994897408 & 253.615292 & -28.176539 & 12.466 & 1.233 & SB1 & Spurious \\
Gaia DR3 6553439603373054720 & 347.377366 & -35.788093 & 12.369 & 14.044 & SB1 & Spurious \\
Gaia DR3 6771307454464848768 & 293.086572 & -23.853788 & 10.435 & 0.748 & SB1 & Spurious \\
\enddata
\end{deluxetable*}

\begin{figure*}
    \centering
    \includegraphics[width=\textwidth]{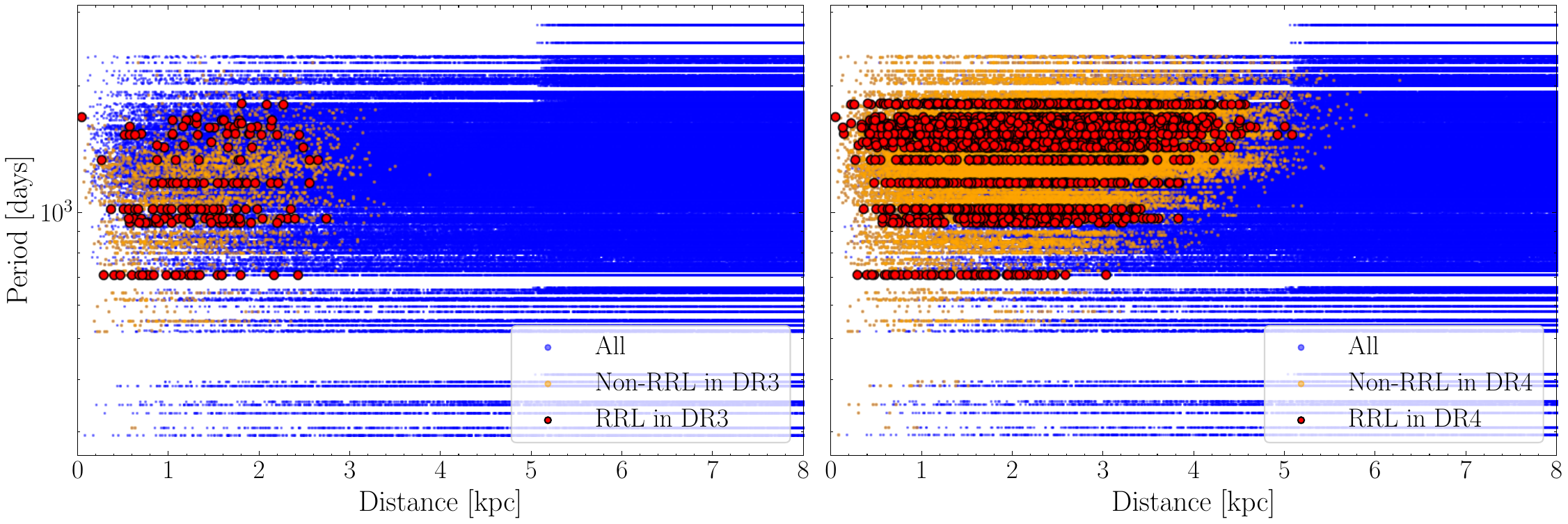}
    \caption{Orbital period vs.\ distance for typical realizations of the Milky Way population of stripped star binaries. Systems detected in DR3 and DR4 are marked by orange circles in the left and right panels respectively. RR Lyrae in binaries that receive orbital solutions are marked with larger red points. As expected, more systems are detectable at larger distances with a longer time baseline. The model predicts that $202^{+20}_{-16}$ and $2655^{+65}_{-15}$ RR Lyrae should receive orbital solutions in DR3 and DR4, respectively. These RR Lyrae have periods spanning $\approx 700$-$1800$ d, with the period distribution peaking at $\approx 1000$ d in DR3 and $\approx 1550$ d in DR4.}
    \label{fig:rrl_fig}
\end{figure*}

We search for RR Lyrae in binaries by cross-matching the \texttt{gaiadr3.vari\_rrlyrae} catalog with the \texttt{gaiadr3.nss\_two\_body\_orbit} and \texttt{gaiadr3.nss\_acceleration\_astro} catalogs. We find 11 candidates, which we list in Table \ref{tab:rrl_binaries}. We discard the EclipsingBinary solutions because the reported periods are aliases of the RR Lyrae pulsation periods. We discard the SB1 solutions because they do not account for stellar variability, implying that the observed radial velocity variation (usually on $\mathcal{O}(1)$ d timescales) is either due to pulsations or strongly contaminated by them. We are left with 2 candidates with astrometric solutions.
Gaia DR3 5239771349915805952, which has an orbital solution, is actually a BY Draconis variable misclassified as an RR Lyrae (as pointed out by \citealt{clementini_rrl_2023}). Gaia DR3 491104466350124544 has a 9-parameter acceleration solution. To vet this system, we retrieve $g$-band and $r$-band light curves from Data Release 23 of the Zwicky Transient Facility (ZTF) survey \citep{bellm_2019}, finding a best-fit period of 0.295594 d from a Lomb-Scargle \citep{Lomb_1976} periodogram analysis. The phase-folded light curves reveal that this system is an eclipsing binary misclassified as an RR Lyrae. Thus, from our investigation, we conclude that no genuine RR Lyrae received astrometric binary solutions in DR3. 

Based on our population synthesis model, we predict that $202^{+20}_{-16}$ RR Lyrae should have received an orbital solution in DR3 (left panel of Figure~\ref{fig:rrl_fig}). These RR Lyrae have periods spanning $\approx 700$-$1800$ d, with the period distribution peaking at $\approx 1000$ d. The RR Lyrae have masses $\approx 0.5\,M_{\odot}$, while their companions range from $0.7\,M_{\odot}$ to $1.8\,M_{\odot}$. Nevertheless, the large majority of these binaries have predicted flux ratios $< 0.1$, since the RR Lyrae are luminous, with absolute magnitudes $M_G \approx 1$. All of these binaries lie within 3 kpc, with the distribution of apparent magnitudes predicted to peak around $G \approx 12$. 

The predicted number of RR Lyrae with orbital solutions in DR3 is in more than $12\sigma$ discrepancy with observations, suggesting that RR Lyrae with au-scale companions are much rarer than the model predicts. In fact, the model predicts a total of 2,696 RR Lyrae formed via binary interactions to fall within 3 kpc (including those without orbital solutions in DR3). \citet{bobrick_iorio_2024} predict that binary-made RR Lyrae from the thin disk should account for $20$--$25$\% of the disk-like RR Lyrae population within 3 kpc. In other words, if the model were accurate, then at least 1 in 5 RR Lyrae (or 539 RR Lyrae) within 3 kpc would be predicted to have a binary companion in an au-scale orbit.

\citet{Giuliano26} perform a complementary population-level analysis of metal-rich RR Lyrae. They begin with a carefully selected observational sample of metal-rich RR Lyrae in the solar neighborhood and investigate whether these systems could be binaries. They sample binary parameters from the model of \citet{bobrick_iorio_2024}, conditioned on the photometric metallicity of each RR Lyrae. They investigate the detectability of RR Lyrae in binaries using a version of \texttt{gaiamock} designed to account for chromaticity biases caused by pulsation (we discuss this bias further in Section~\ref{sec:rrl}). Based on their fiducial simulations, they predict fewer RR Lyrae in binaries to receive astrometric binary solutions in DR3 because (a) the observed sample of RR Lyrae is not complete, (b) the \citet{bobrick_iorio_2024} model predicts a higher space density of metal-rich RR Lyrae than implied by the observed sample, and (c) photometric variability reduced the fraction of RR Lyrae that receive orbital solutions. \citet{Giuliano26} nevertheless rule out large binary fractions at high confidence, unless the binary periods are significantly shorter or longer than predicted by \citet{bobrick_iorio_2024}. This is broadly consistent with our results.


\subsection{Black hole impostors} 

\begin{figure*}
    \centering
    \includegraphics[width=\textwidth]{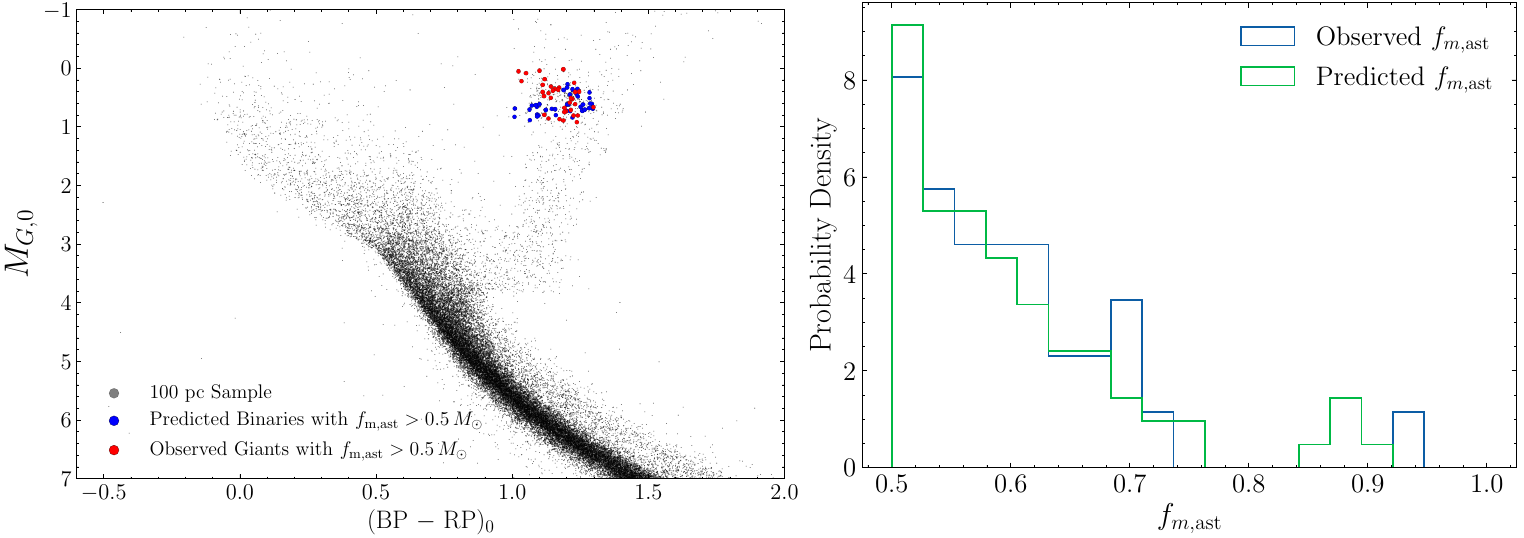}
    \caption{Left: Extinction-corrected color-magnitude diagrams (CMD) of red clump stripped star binaries predicted to have high astrometric mass functions $f_{m, \text{ast}} > 0.5$, assuming full rejuvenation of their companions following mass transfer. We plot a random sample of sources from \textit{Gaia} DR3 within 100 pc for comparison. Right: Normalized histograms of astrometric mass functions of these stripped star binaries. The model reproduces the observed sample of red clump stars with DR3 orbital solutions implying $f_{m, \text{ast}} > 0.5$. These binaries are ``BH impostors," since their orbits imply $\gtrsim 3\,M_{\odot}$ dark companions when they are (mis)interpreted as systems without a history of mass transfer.}
    \label{fig:rc_cmd_figure}
\end{figure*}

\textit{Gaia} DR3 features a population of red giant and red clump stars with relatively large astrometric mass functions, implying massive companions (left panel of Figure~\ref{fig:rc_cmd_figure}). Here, the astrometric mass function $f_{m,\text{ ast}} \equiv (a_0^3 \times 365.25^2) / (P^2 \varpi^3)$, where $a_0$ is the photocenter semi-major axis, $P$ is the orbital period, and $\varpi$ is the parallax. We define the red clump as the group of stars with $1.0 > M_{G, 0} > 0.0$ and $1.3 > (\text{BP} - \text{RP})_0 > 1.0$. 

Since red clump stars had larger radii in the past, these binaries likely had a history of mass transfer. Hence, these systems could host a partially stripped giant orbiting a fainter, spun-up luminous companion. Since this companion would have broadened spectral lines, it could evade detection in the optical \citep[e.g.,][]{bodensteiner_hr_2020, shenar_hidden_2020}. Furthermore, mass transfer can cause the less massive, stripped star in a binary to be the more luminous and evolved component. In these cases, the mass of the luminous stripped star can be overestimated, leading to the erroneous conclusion that the unseen companion is a black hole (BH) \citep[e.g.,][]{el-badry_ngc_2022, el-badry_what_2022, el-badry_unicorns_2022}. Indeed, while there have been several ongoing attempts to find dormant BHs in wide binaries in the past few decades, extensive spectroscopic searches have only recently yielded reliable detections of dormant BHs \citep[e.g.,][]{giesers_detached_2018, shenar_x-ray-quiet_2022}. This is because BH binaries are relatively rare, while ``BH impostors" like the aforementioned stripped star binaries are comparatively abundant.

Recently, \citet{wang_gap_2024} reported the discovery of Gaia DR3 3425577610762832384 (hereafter G3425), a binary hosting a red giant orbiting a candidate mass-gap BH in a wide, 880-d circular orbit. They inferred a companion mass of $3.6^{+0.8}_{-0.5}\,M_{\odot}$, assuming a spectroscopic mass for the red giant of $2.66^{+1.18}_{-0.68}\,M_{\odot}$. Since few BHs are known with masses in the range of $3$--$5\,M_{\odot}$, and low-mass BHs are predicted to be born with large natal kicks that result in eccentric orbits \citep[e.g.,][]{burrows_theory_2024}, this system stands out from the known BH population and challenges standard evolutionary models. \citet{wang_gap_2024} constrained the mass of the unseen component using a combination of LAMOST radial velocities and Gaia DR2 and DR3 astrometry, and performed spectral disentangling on the low-resolution LAMOST data to try and rule out the presence of a luminous secondary. In doing so, \citet{wang_gap_2024} assumed that the source is a first-ascent red giant. However, it falls directly on top of the red clump in the color-magnitude diagram, suggesting that it could instead by a helium-burning star that was recently stripped by mass transfer. In this case, its mass might be significantly lower than estimated with single-star models.

G3425 does not have an astrometric solution, so its mass function is uncertain. However, there are several similar sources that do have orbital solutions. To assess whether they could be mistaken for mass-gap BH binaries, we investigate the systems with the highest astrometric mass functions in the simulated population of \citet{bobrick_iorio_2024} that lie on or near the red clump on an extinction-corrected color-magnitude diagram. We first adopt the $G$-band magnitudes predicted by \citet{bobrick_iorio_2024} for the stripped stars' companions. Most of these companions are somewhat evolved --- i.e., they are brighter than they would be at the zero-age main sequence (ZAMS). In this case, the model produces no sources with $f_{m,\text{ ast}} > 0.5\,M_{\odot}$ and cannot explain the observed tail of high-$f_{m,\text{ ast}}$ binaries. However, it is possible that rotationally-limited accretion underestimates the efficiency of mass transfer \citep[e.g.,][]{lechien_efficient_2025}. In this case, the companion star may be rejuvenated more than assumed in the calculations of \citet{bobrick_iorio_2024}, moving it toward the main sequence. This, in turn, would increase the observed astrometric mass function at the same mass ratio.

Motivated by this, we next consider the limit of full rejuvenation, which returns all companions to the ZAMS for their post-accretion mass. We re-compute companion $G$-band magnitudes and flux ratios based on MIST models for main sequence stars \citep{choi_2016}. For simplicity, we adopt the default assumptions of solar metallicity and $v/v_{\text{crit}} = 0.4$, though varying composition and rotation does not significantly change our results. Then, we re-run \texttt{gaiamock} to predict astrometric mass functions and retrieve all stripped star binaries in the red clump with $f_{m, \text{ ast}} > 0.5$. We find that these binaries host $\approx 0.5\,M_{\odot}$ horizontal branch stars that outshine their $0.6$--$1.5\,M_{\odot}$ main sequence companions. If the masses of these stripped giants are overestimated due to their somewhat blue unresolved colors, then these systems represent the population of BH impostors that could be mistaken for binaries hosting mass-gap BH companions.

We compare the distributions of predicted and observed astrometric mass functions in the right panel of Figure~\ref{fig:rc_cmd_figure}. We find that, in the limit of full rejuvenation, the model of \citet{bobrick_iorio_2024} is able to reproduce the observed population of red clump stars with $f_{m, \text{ast}} > 0.5$. Specifically, the normalized histograms are similar for both the simulated and observed populations. While we find that the raw counts in the simulated population are overestimated by a factor of $\sim 2$ relative to the observed population, this is unsurprising, given the overestimation factors for other classes of stripped stars. These objects have DR3 orbital solutions implying massive unseen companions, but no obvious sign of a luminous companion to the red giant star. We defer a more detailed investigation of the observed sample, including follow-up spectroscopy, to future work.

\subsection{Looking ahead to DR4}

\begin{figure*}
    \centering
    \includegraphics[width=0.85\textwidth]{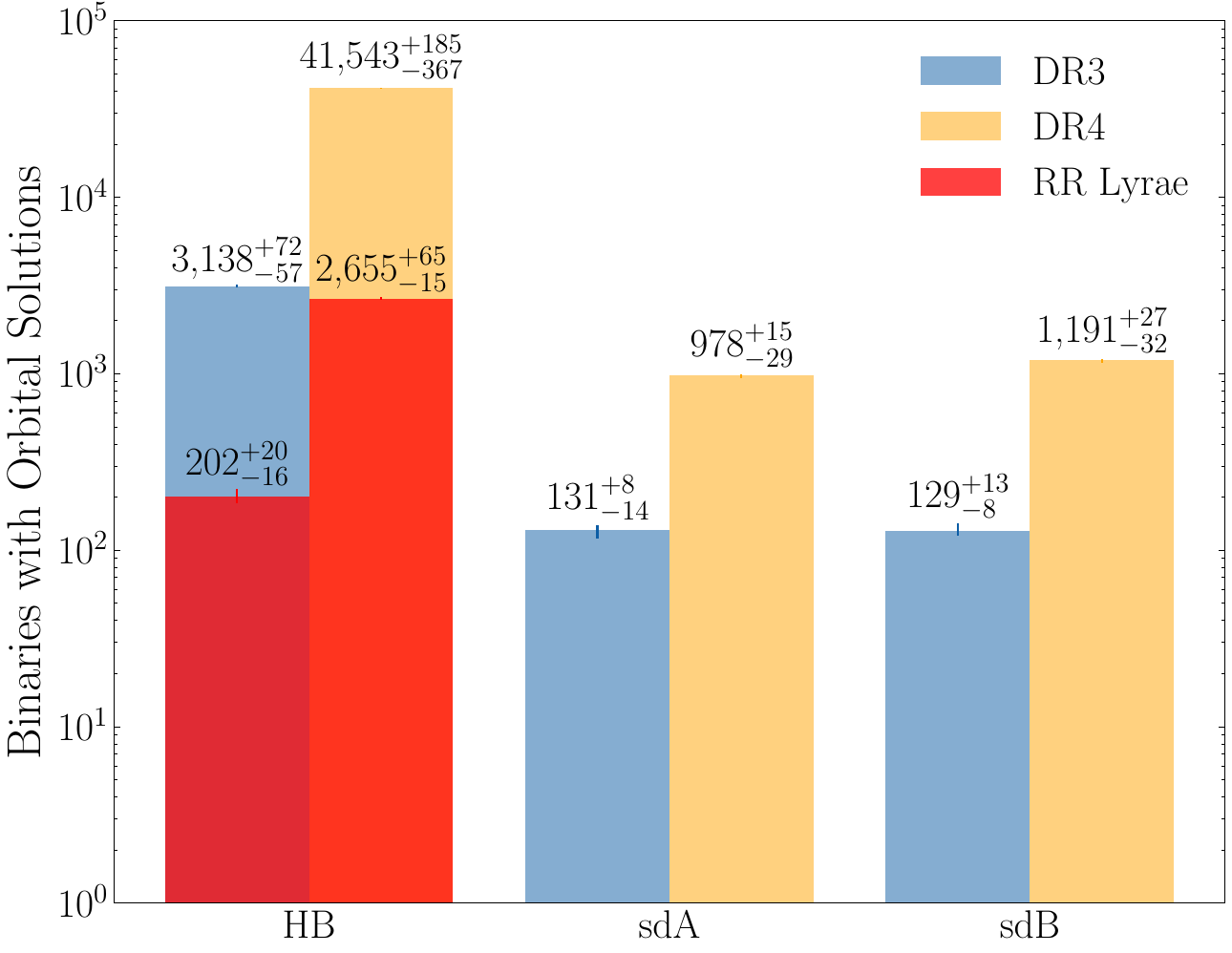}
    \caption{Median number of detected stripped star binaries in DR3 and DR4 across 10 realizations, broken down by class. The error bars signify the middle 68\% of counts. A longer time baseline leads to a stark increase in the number of stripped star binaries that receive orbital solutions, with the number of detected binaries hosting RR Lyrae being comparable to the number of detected binaries hosting sdA or sdB stars.}
    \label{fig:class_bar_plot}
\end{figure*}

\begin{figure*}
    \centering
    \includegraphics[width=\textwidth]{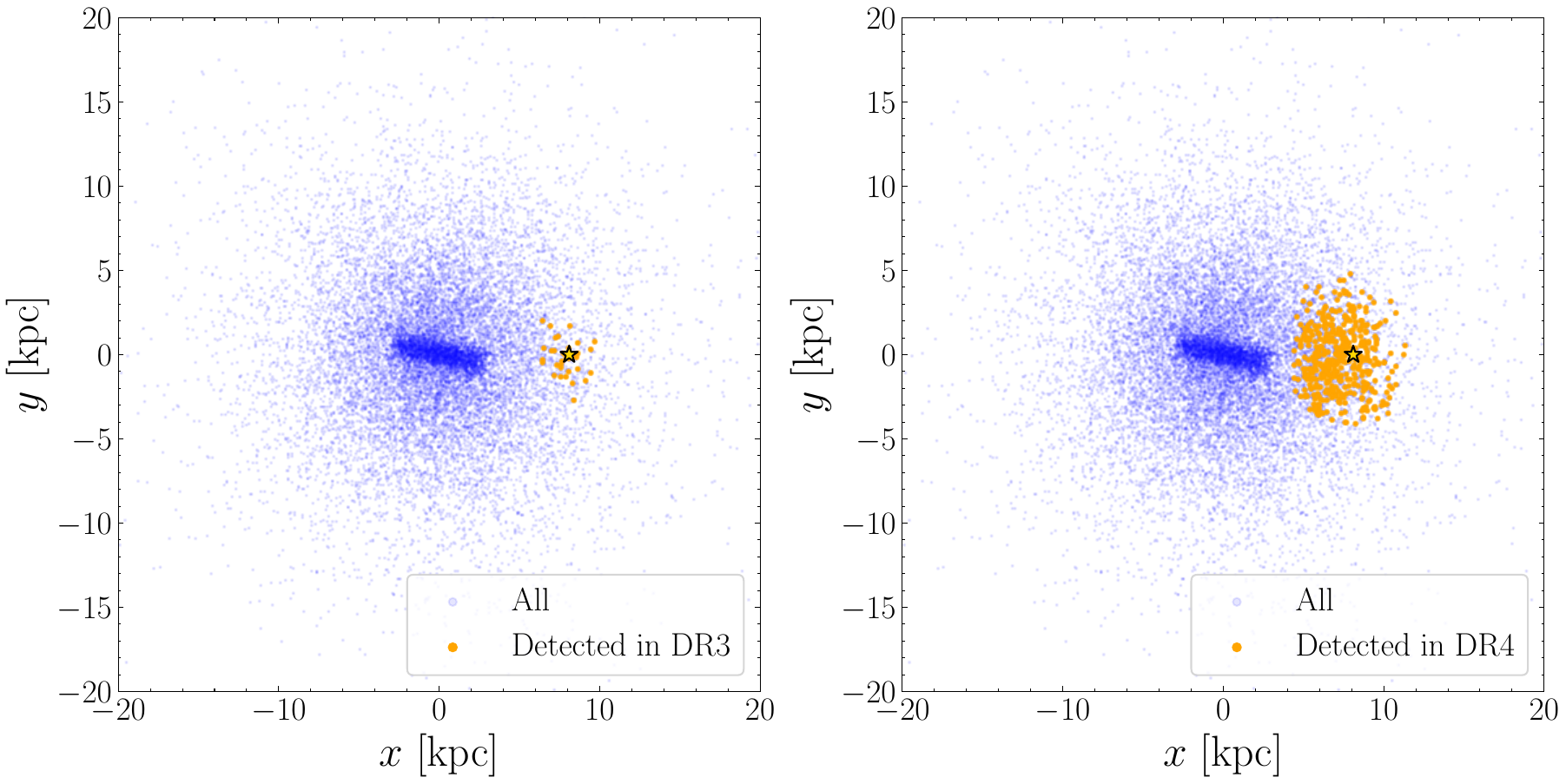}
    \caption{Predicted spatial distribution of stripped stars in binaries. We show all systems in blue, and the systems detected in either DR3 or DR4 (for a typical MW realization) in orange, plotting 1\% of each set of binaries for clarity. The detected binaries are clustered around the location of the Sun, with a longer time baseline increasing the volume to which \textit{Gaia} is sensitive.}
    \label{fig:mw_dist_figure}
\end{figure*}

\begin{figure*}
    \centering
    \includegraphics[width=\textwidth]{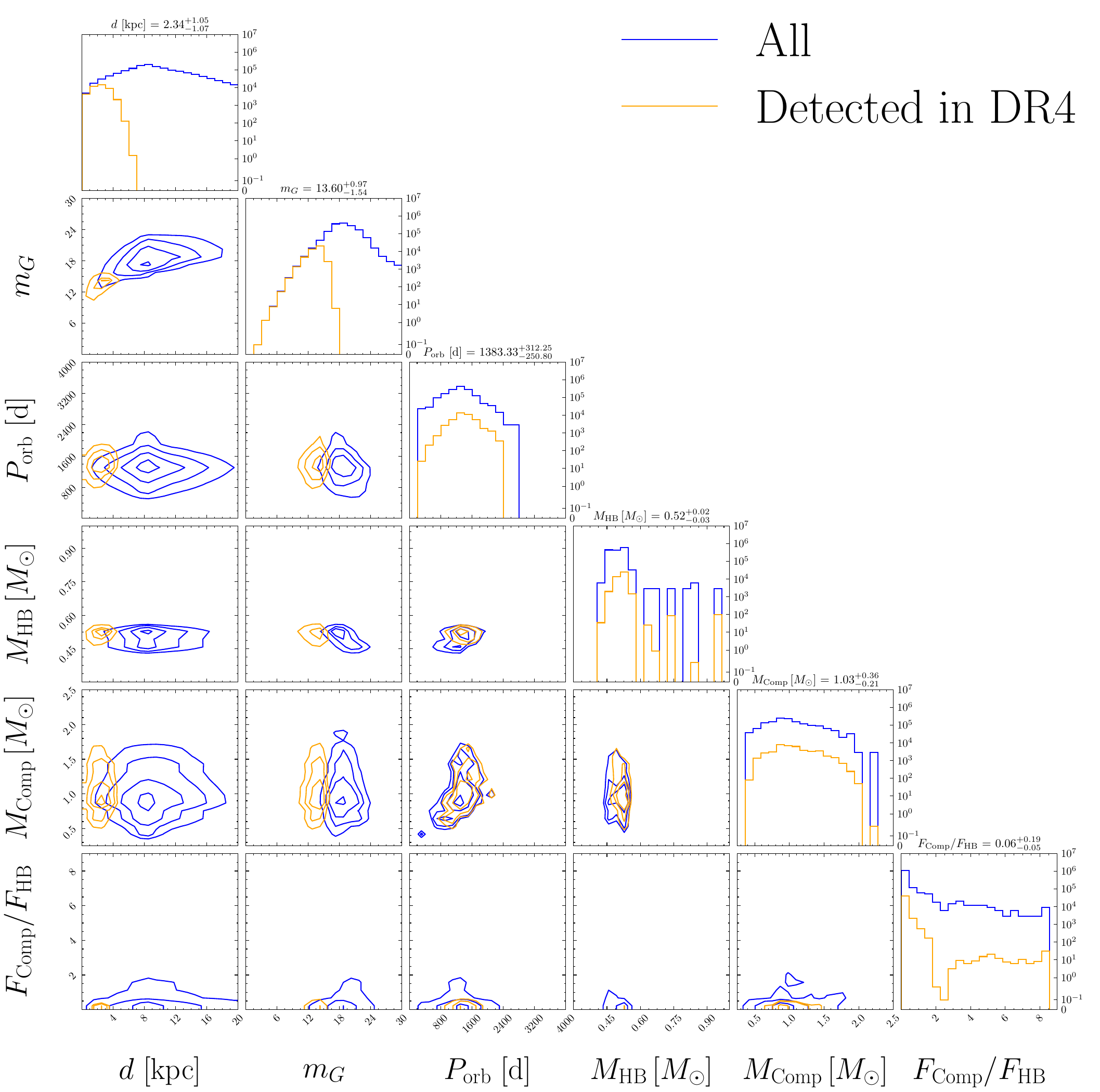}
    \caption{Properties of all stripped star binaries predicted by the model (blue) and those predicted to be detected in DR4 (orange). The diagonal entries display the marginal distribution of each parameter, while each of the other panels displays a joint distribution. Contours identify regions of high density in each parameter space. To reduce shot noise, we combine predictions from 10 realizations, each corresponding to a different placement of the Sun in the simulated galaxy. We then rescale the marginal distributions by 1/10, so that the counts show the average number of binaries per realization. We find that distance is the most important variable; all stripped star binaries that are nearby receive orbital solutions, but no binaries beyond 7 kpc are detected. Furthermore, all binaries brighter than $G \approx 13$ receive orbital solutions, but no binaries fainter than $G \approx 18$ being detected. The predicted distribution of flux ratios peaks at $\lesssim 0.1$, and displays a gap at around $\approx 2$, where the predicted photocenter orbits are small.}
    \label{fig:dr4_corner_plot}
\end{figure*}

We show the median number of detected stripped star binaries across 10 realizations in DR3 and DR4 in Figure~\ref{fig:class_bar_plot}, with the error bars showing the middle 68\% of counts. We break detections down by class. Assuming the same cuts and astrometric cascade in DR4 as in DR3, the adopted model predicts that $1,191^{+27}_{-32}$ sdB binaries, $978^{+15}_{-29}$ sdA binaries, and $41,453^{+185}_{-367}$ HB binaries will receive astrometric orbital solutions in DR4. Among the HB binaries, $2,655^{+65}_{-15}$ systems are expected to host metal-rich RR Lyrae. We note that these are likely overestimates, since the model overpredicts all classes in DR3. Calibrating to the observed sample of sdBs in wide orbits with orbital solutions in DR3, we predict that $\approx 157 \pm 4$ sdB binaries will be detected in DR4.

We plot orbital period versus distance for typical realizations of the Galactic population of stripped star binaries in Figure~\ref{fig:rrl_fig}. We mark systems detected in DR3 and DR4 with orange circles in the left and right panels, respectively. RRL in binaries that receive astrometric binary solutions are marked with larger red points. As expected, a longer time baseline leads to more systems being detected at farther distances, with $202^{+20}_{-16}$ and $2655^{+65}_{-15}$ RRL receiving orbital solutions in DR3 and DR4, respectively. These detected RRL have periods spanning $\approx 700$-$1800$ d, with the period distribution peaking at $\approx 1000$ d in DR3 and $\approx 1550$ d in DR4.

We show the Galactic distribution in Cartesian coordinates of the synthetic population of stripped star binaries in Figure~\ref{fig:mw_dist_figure}. We plot the entire population in blue, and the subset of detected systems in orange, with the left and right panels displaying our predictions for DR3 and DR4, respectively. We plot 1\% of each set of binaries for clarity. As expected, we find that the detected binaries are clustered around the Sun, with \textit{Gaia} being sensitive to a much larger volume of binaries in DR4 than in DR3.

Figure~\ref{fig:dr4_corner_plot} summarizes the properties of the detected and full stripped star populations predicted by the model. We combine the predictions from all 10 realizations, and then rescale all marginal distributions so that the counts show the average number of binaries per realization. We plot the distributions for the entire population and detected subset in blue and orange, respectively. We find that the detected binaries are closer and brighter on average than the underlying population, but display a similar range of orbital periods, component masses, and flux ratios. We find that distance is the most important parameter, with all nearby binaries receiving orbital solutions, but no binaries being detected beyond 7 kpc. All binaries brighter than $G \approx 13$ receive orbital solutions, with no binaries fainter than $G \approx 18$ being detected. The predicted distribution of flux ratios peaks at $F_{\text{Comp}} / F_{\text{HB}} \lesssim 0.1$, and displays a gap at a flux ratio of $\approx 2$, where the photocenter orbit is small (see the right panel of Figure~\ref{fig:mock_orbits}).

\section{Discussion} 
\label{sec:discussion}

\subsection{Implications for the Galactic population of stripped star binaries}

\subsubsection{Overestimation of stripped stars within 500 pc}

Based on the color-magnitude diagram cuts proposed by \citet{vos_bobrick_2020}, the model of \citet{bobrick_iorio_2024} predicts $\approx 2\times$ more composite sdB binaries within 500 pc than observed in the sample of \citet{dawson_volume_2024}, which is reported to be 90\% complete. This conclusion holds even after requiring that the unresolved binary falls in the sdB clump of the color-magnitude diagram, and that the companion produces detectable infrared excess. 

Other binary population synthesis (BPS) models also overestimate the sdB birth rate. For example, \citet{dawson_volume_2024} found that the population synthesis model of \citet{han_origin_2003} produced an sdB birth rate $\sim 15$--$30\times$ higher than the birth rate implied by their volume-limited observational sample. However, \citet{han_origin_2003} do not use a Besan\c{c}on model, and overestimate the Galactic star formation rate by a factor of $\sim 10$ \citep{vos_bobrick_2020}. \citet{rodriguez_ruiter_2025} compared observed populations of sdBs \citep{culpan_subdwarf_2022, dawson_volume_2024} against a simulated Galactic population generated using the COMPAS rapid BPS code \citep{compas_2022}. They investigated hot subdwarfs formed both via common-envelope evolution and stable mass transfer, but did not perform detailed \texttt{MESA} simulations. They found that their model predicts a large number of sdBs with early-type companions $> 1\,M_{\odot}$, which observational samples are biased against. While removing these systems improved the agreement between the observed and theoretical populations, they concluded that the predicted number of sdB binaries within 500 pc is still too large even after this change. They cite uncertainties in the star formation rates and local densities assumed within their implementation of the Besan\c{c}on model as a potential explanation for this overprediction.  

\subsubsection{Overestimation of stripped star binaries with DR3 astrometric binary solutions}

The model of \citet{bobrick_iorio_2024} overestimates the number of composite hot subdwarf binaries that receive astrometric orbital solutions in DR3 by an factor of $\approx 5.5$. When restricted to composite sdBs within 500 pc with DR3 orbital solutions, the overestimation factor rises to $\approx 19$. Observational incompleteness due to shot noise could be one potential explanation. If the volume-limited sample of \citet{dawson_volume_2024} missed $2$--$3$ composite sdBs within 500 pc that received orbital solutions in DR3, then these overestimation factors would become consistent with each other. Another explanation could be that the flux ratios and orbital periods predicted by the model are different from those that are observed. Indeed, in the adopted model, the median $G$-band flux ratio is predicted to be 0.432. Meanwhile, in the observed sample of sdBs with long-period spectroscopic orbits, the median $G$-band flux ratio is $\approx 1.5$ \citep[e.g.,][]{barlow_two_2013, molina_wide_2026}. The lower median flux ratio explains why the sdB binaries in the simulated population are easier to detect with DR3 astrometry than the observed composite sdB binaries (e.g. Figure~\ref{fig:mock_orbits}). The model specifically predicts too many sdBs with low-mass companions ($\lesssim 0.7\,M_{\odot}$) and wide orbits, which are easily detectable with astrometry but not spectroscopically composite. Perhaps the simplest explanation is that mass transfer in these systems is less stable than assumed in the model, such that sdBs with faint and low-mass companions become post common envelope systems with short periods. 

The overestimation factor for composite sdB binaries with 7-parameter acceleration solutions in DR3 is a factor of $\approx 2$ higher than the corresponding factor for orbital solutions, lending further credence to the hypothesis that there is a difference in the predicted and observed parameters of sdB binaries. In the adopted model, the median orbital period of the sdB binaries is predicted to be 1107 d. On the other hand, in the observed sample of sdBs with long-period spectroscopic orbits, the median orbital period is 963 d \citep[e.g.,][]{barlow_two_2013, otani_orbital_2018, molina_wide_2026}. Moreover, the median orbital period of the sdBs with orbital solutions in DR3 is just 821 d. We conclude that there are fewer acceleration solutions relative to orbital solutions, and fewer astrometric solutions relative to spectroscopic solutions, in the observed population because the model is predicting, on average, orbital periods that are longer than observed.

\subsubsection{Overestimation of binary RR Lyrae}
\label{sec:rrl}

The results of this study also provide strong constraints on the formation channels of partially stripped stars in binaries. Indeed, because the predicted periods of the RR Lyrae in binaries are well-matched to \textit{Gaia}, the limits we can set here are much more stringent than those that could be set using RV follow-up studies, eclipses, or any other search for binary RR Lyrae. We find that over $200$ metal-rich RR Lyrae are predicted to receive orbital solutions in DR3, with that number increasing by a factor of $\approx 10$ in DR4. Hence, the fact that no such RR Lyrae are actually observed in DR3 suggests that young, metal-rich RR Lyrae do not efficiently form in binaries, at least under the assumptions of the \citet{bobrick_iorio_2024} model. 

One possible explanation for the discrepancy with observations is a variability bias that reduces the probability that RR Lyrae receive orbital solutions in DR3. Indeed, the color of a RR Lyrae variable varies as its pulsates, and \textit{Gaia}'s astrometric calibration accounts for chromaticity \citep{lindegren_gaia_2021}. \citet{Giuliano26} develop a model to approximately account for effects of variability within the \texttt{gaiamock} framework. We tested this model on our simulated population and find that compared to the fiducial version of  \texttt{gaiamock}, it predicts about half the number of RR Lyrae to receive astrometric orbits, such that $\approx 100$ should receive orbits in DR3. 
Thus, accounting for variability bias likely reduces but does not solve the tension between the model predictions and DR3 data. 

\subsection{Explaining the difference between predictions and observations}

Differences between the simulated and observed populations can be explained by any combination of the following:

\begin{enumerate}
    \item the proposed binary formation channel for stripped stars is incorrect, 
    \item the assumptions of the adopted binary population synthesis model need to be modified,
    \item or there is an observational bias that is yet unaccounted for.
\end{enumerate}

We now consider these in turn.

\subsubsection{The proposed binary formation channel for stripped stars}

The simplest explanation for the tension between the predicted and observed stripped star populations is that the proposed binary formation channel is incorrect. For example, it could be that mass transfer is less stable than the model assumes, leading to fewer long-period sdBs being produced than expected. Another possibility is that the model where RR Lyrae form via partial envelope stripping is inefficient in nature, perhaps because mass transfer usually results in near-complete stripping once it begins. 

This does not necessarily imply that metal-rich RR Lyrae cannot form via binary channels, since other models may exist for forming RR Lyrae via binary evolution. A notable example is the pathway investigated by \citet{rui_fuller_2024}, in which a merger between a main sequence star and a helium white dwarf (i.e., a cataclysmic variable-like system) produces a red giant branch remnant that later evolves into a core helium-burning star. \citet{rui_fuller_2024} point out that low-mass remnants can produce horizontal branch stars or hot subdwarfs. If a horizontal branch star descending from such a merger remnant falls in the instability strip, it could be observable as a single RR Lyrae today.

\subsubsection{Assumptions of the binary population synthesis model}

Significant uncertainties remain in the input binary physics of the adopted population synthesis model. These include the assumed orbital evolution during mass transfer. Indeed, the final orbital period distribution, which strongly influences detectability with \textit{Gaia}, is quite sensitive to assumptions about binary mass transfer and angular momentum loss. Though sdBs in wide orbits are generally thought to form via highly non-conservative mass transfer \citep[e.g.,][]{vos_bobrick_2020}, recent work has shown that the mass transfer in hot subdwarf binaries could be more conservative than previously thought \citep[e.g.,][]{lechien_efficient_2025}.\footnote{\citet{lechien_efficient_2025} analyze a sample of Be+sdOB binaries, which are produced by more massive progenitors than those considered in this work. Nevertheless, it is still possible that mass transfer in sdB binaries formed from low-mass progenitors is more conservative than previously assumed.} If mass transfer were more efficient, the predicted flux ratios would be higher, and fewer hot subdwarfs would be likely to get orbital solutions. Furthermore, observational studies have shown that stripped star binaries at longer orbital periods tend to have larger eccentricities, up to $e \sim 0.25$ at $P \sim 1350$ d \citep[e.g.,][]{vos_eccentricity_2015, vos_orbits_2017}. Since \textit{Gaia} is less sensitive to eccentric orbits \citep{el-badry_generative_2024}, the assumption of tidal circularization could lead to an overprediction in the number of stripped star binaries detectable in DR3. Finally, modifications to the mass transfer prescription could cause binary-made RR Lyrae to end up with longer orbital periods than the model predicts. In that case, fewer RR Lyrae variables would be detectable in DR3, ameliorating the tension with observations.

\subsubsection{Observational biases}

Our modeling of observational selection effects accounts for biases due to dust extinction and the spatial distribution of binaries in the Galactic model. However, variability may introduce a bias against the detection of RR Lyrae in binaries, since the color index of these stars varies as they pulsate, and the \textit{Gaia} along-scan measurements depend on measured $\mathrm{BP}-\mathrm{RP}$ color \citep{lindegren_gaia_2021}. We discuss the impact of this effect briefly in Section~\ref{sec:rrl}, with a deeper investigation forthcoming in \citet{Giuliano26}. Our investigation suggests that observational biases caused by variability alone cannot explain the lack of RR Lyrae with orbital solutions in DR3, though variability may reduce the number of accepted astrometric solutions by a factor of $\approx 2$.

While the astrometric cascade implemented in \texttt{gaiamock} is somewhat simplified compared to the one actually used in constructing the published DR3 binary catalogs, \citet{el-badry_generative_2024} show that their forward model can be used to generate a mock catalog of astrometric binaries quite similar to the catalog actually published in DR3. While they find that their mock catalog contains more high-eccentricity orbits than observed, \textit{Gaia}'s sensitivity only falls steeply above $e \gtrsim 0.5$ \citep[e.g.,][]{wu_eccentricities_2025}. Hence, sdB binaries would have to be quite eccentric to remain undetected; in contrast, all interacting binaries in the simulated population of \citet{bobrick_iorio_2024} are assumed to have tidally circularized as a result of mass transfer. If there is an unidentified bias preventing RR Lyrae or sdBs from receiving astrometric orbits, it thus would have to affect them differently from ordinary binaries.\footnote{Section~\ref{sec:comparison} suggests that \texttt{gaiamock} might underestimate the scatter in parallax errors at fixed magnitude. However, such a bias would only lead to an underestimation of the number of wide stripped star binaries that receive astrometric orbital solutions.}

\subsubsection{Explaining black hole impostors}

When mass transfer causes the less massive component of a stripped star binary to become the more luminous primary, its mass can be overestimated by single-star models, causing the binary to be mistaken for a system hosting a dormant BH. We find that stripped He-burning giants provide a reasonable explanation for the giants with high observed astrometric mass functions in DR3, although some tweaking of the mass transfer efficiency is required to match the data. Indeed, for the adopted model to be able to reproduce the observed sample of red clump stars with $f_{m,\text{ ast}} > 0.5\,M_{\odot}$ in DR3, we require mass transfer to fully rejuvenate the companions to those red clump stars, returning those companions to the main sequence.\footnote{Applying this modified prescription to the RR Lyrae binaries would not significantly change their detectability, since giants on the horizontal branch stars are expected to outshine their companions. On the other hand, applying this modified prescription to the hot subdwarf binaries would enhance their detectability by reducing the median flux ratio of that population. Nevertheless, we do not investigate this further, since it is unclear whether the same mass transfer physics can be assumed regardless of the degree of stripping.} This suggests that a deeper investigation into the mass transfer physics and accretor evolution assumed in BPS modeling is warranted. 

\section{Conclusion}
\label{sec:conclusion}

We have applied the generative model of \citet{el-badry_generative_2024} to a simulated Galactic population of low-mass stripped core helium-burning stars with main sequence or red giant companions \citep{bobrick_iorio_2024} to predict the number of these systems that will receive astrometric binary solutions in \textit{Gaia} DR3 and DR4. The stripped stars include red clump stars, horizontal branch (HB) stars (a fraction of which are RR Lyrae variables), and hot A-type (sdA) or B-type (sdB) subdwarfs. We are already able to test the validity of this model by comparison to observations in DR3; furthermore, we will be able to assess the model's soundness even more vigorously in DR4. We summarize our results below.

\begin{itemize}
    \item By comparison to the volume-limited sample of \citet{dawson_volume_2024}, we find that the adopted model reproduces the number of composite sdB systems within 500 pc to within a factor of $\approx 2$ (Figure~\ref{fig:obs_bar_plot}). However, based on observations of known sdBs and sdB candidates in the \textit{Gaia} DR2 and EDR3 samples of \citet{geier_dr2_2019} and \citet{culpan_subdwarf_2022} (Table~\ref{tab:sdB_sols}), we find that the adopted model strongly overpredicts the number of hot subdwarfs that receive astrometric binary solutions (Figure~\ref{fig:obs_bar_plot}), with the overestimation factor being larger for acceleration solutions than orbital solutions. Based on simulated detection probabilities, we deduce that this discrepancy can be partially explained by a mismatch between the flux ratios of the simulated and observed sdB populations (Figures~\ref{fig:vos_cmd} and \ref{fig:a0_comp_plot}).
    \item The model predicts that many young, metal-rich RR Lyrae in wide binaries should be detectable in DR3 and DR4 (Figures~\ref{fig:rrl_fig} and \ref{fig:class_bar_plot}). However, no such RR Lyrae binaries have been found in DR3 (Table~\ref{tab:rrl_binaries}). This implies that metal-rich RR Lyrae do not form efficiently via the binary channel modeled by \citet{bobrick_iorio_2024}, or that pulsational variability introduces a bias against the detection of binary RR Lyrae in DR3. 
    
    Starting from a carefully selected observational sample of metal-rich RR Lyrae, \citet{Giuliano26} perform a complementary population-level analysis of RR Lyrae in wide binaries, implementing a forward model for \textit{Gaia} detectability designed to account for chromaticity biases caused by pulsation. Based on a Bayesian inference framework, they use the lack of detections in DR3 to rule out large RR Lyrae binary fractions at high confidence across most of the metallicity range, in broad agreement with our results.

    \item \textit{Gaia} DR3 features a population of red clump stars with relatively large astrometric mass functions, implying massive companions (Figure~\ref{fig:rc_cmd_figure}). If the companion $G$-band magnitudes predicted by \citet{bobrick_iorio_2024} are adopted, the model produces no sources with $f_{m,\text{ ast}} > 0.5\,M_{\odot}$ and cannot explain this observed tail of high-$f_{m,\text{ ast}}$ binaries. This is because many of the partially stripped star binaries that fall in the red clump region of the color-magnitude diagram have evolved companions and relatively large flux ratios. On the other hand, in the limit of full rejuvenation (which assumes that all companions are returned to the zero-age main sequence for their post-accretion mass), the model is able to reproduce the observed population of red clump stars with $f_{m, \text{ast}} > 0.5$ (Figure~\ref{fig:rc_cmd_figure}). We conclude that stripped He-burning giants provide a reasonable explanation for the ``BH impostors'' observed in DR3, although modification of the stellar response to accretion is required to match the data.
    \item Assuming the same cuts and astrometric cascade in DR4 as in DR3, the adopted model predicts that $1,191^{+27}_{-32}$ sdB binaries, $978^{+15}_{-29}$ sdA binaries, and $41,453^{+185}_{-367}$ HB binaries will be detectable in DR4 (Figure~\ref{fig:class_bar_plot}). These are likely overestimates, since the model overpredicts all classes in DR3. Nevertheless, these binaries are predicted to have orbital periods of several hundred to a few thousand days, and are clustered around the location of the Sun in the MW-like galaxy (Figure~\ref{fig:mw_dist_figure}). They are closer and brighter than the underlying population, with predicted secondary-to-primary flux ratios $\lesssim 0.1$ (Figure~\ref{fig:dr4_corner_plot}). 
\end{itemize}

DR4's sensitivity to longer-period orbits will more strongly test the predictions of binary population synthesis models for stripped stars in the wide orbit regime.


\begin{acknowledgments}
We thank the referee for useful comments. This research was supported by NSF grants AST-2540180 and AST-2307232. This research was supported in part by grant NSF PHY-2309135 to the Kavli Institute for Theoretical Physics (KITP). AB acknowledges support from the Australian Research Council (ARC) Centre of Excellence for Gravitational Wave Discovery (OzGrav), through project number CE230100016. GI is supported by a fellowship grant from
la Caixa Foundation (ID 100010434). The fellowship code is
LCF/BQ/PI24/12040020. This work has made use of data from the European Space Agency (ESA) mission
{\it Gaia} (\url{https://www.cosmos.esa.int/gaia}), processed by the {\it Gaia}
Data Processing and Analysis Consortium (DPAC,
\url{https://www.cosmos.esa.int/web/gaia/dpac/consortium}). Funding for the DPAC
has been provided by national institutions, in particular the institutions
participating in the {\it Gaia} Multilateral Agreement. 
\end{acknowledgments}

\begin{contribution}

PN led the analysis and writing of the paper. KE came up with the initial research concept and obtained the funding. AB supplied the binary population synthesis modeling. GI offered insight on the detectability of metal-rich RR Lyrae in binaries. FM, JV, and MV provided comments on the observed sample of hot subdwarfs in wide binaries.


\end{contribution}

%

\software{astropy \citep{2013A&A...558A..33A, 2018AJ....156..123A, 2022ApJ...935..167A}}





\bibliography{bibliography}{}
\bibliographystyle{aasjournalv7_modified}



\end{document}